\documentclass[lettersize,journal]{IEEEtran}
\usepackage{amsmath,amsfonts}
\usepackage{algorithmic}
\usepackage{algorithm}
\usepackage{array}
\usepackage[caption=false,font=small,labelfont=sf,textfont=sf]{subfig}
\usepackage{textcomp}
\usepackage{stfloats}
\usepackage{url}
\usepackage{verbatim}
\usepackage{graphicx}
\usepackage{cite}
\usepackage{amssymb}
\usepackage{xcolor} 
\usepackage{hyperref}
\usepackage{tabularx}
\usepackage{booktabs}
\usepackage{multirow}
\hypersetup{hypertex=true,
	colorlinks=true,
	linkcolor=blue,
	anchorcolor=blue,
	citecolor=blue,
	urlcolor=black}

\begin{document}

\title{Generative Learning Enhanced Intelligent Resource Management for Cell-Free Delay Deterministic Communications}

\author{Shuangbo Xiong, Cheng Zhang,~\IEEEmembership{Member,~IEEE}, Wen Wang, Wenwu Yu,~\IEEEmembership{Senior Member,~IEEE}, and Yongming Huang,~\IEEEmembership{Fellow,~IEEE}
\thanks{S. Xiong, C. Zhang, and Y. Huang are with the National Mobile Communications Research Laboratory, Southeast University, Nanjing 210096, China, and Purple Mountain Laboratories, Nanjing 211111, China (email: shuangbo\_xiong@seu.edu.cn, zhangcheng\_seu@seu.edu.cn, huangym@seu.edu.cn). (Corresponding author: C. Zhang).}
\thanks{W. Wang is with the Purple Mountain Laboratories, Nanjing 211111, China (email: wangwen@pmlabs.com.cn).}
\thanks{W. Yu is with the School of Mathematics, Frontiers Science Center for Mobile Information Communication and Security, Southeast University, Nanjing 210096, and also with the Purple Mountain Laboratories, Nanjing 211111, China (e-mail: wwyu@seu.edu.cn).}
}

\markboth{Journal of \LaTeX\ Class Files,~Vol.~14, No.~8, August~2021}%
{Shell \MakeLowercase{\textit{et al.}}: A Sample Article Using IEEEtran.cls for IEEE Journals}


\maketitle

\begin{abstract}
Cell-free multiple-input multiple-output (CF-MIMO) architecture significantly enhances wireless network performance, offering a promising solution for delay-sensitive applications. This paper investigates the resource allocation problem in CF-MIMO systems, aiming to maximize energy efficiency (EE) while satisfying delay violation rate constraint. We design a Proximal Policy Optimization (PPO) with a primal-dual method to solve it.
To address the low sample efficiency and safety risks caused by cold-start of the designed safe deep reinforcement learning (DRL) method, we propose a novel offline pretraining framework based on virtual constrained Markov decision process (CMDP) modeling.
The virtual CMDP consists of reward and cost prediction module, initial-state distribution module and state transition module. Notably, we propose an evidence-aware conditional Gaussian Mixture Model (EA-CGMM) inference approach to mitigate data sparsity and distribution drift issues in state transition modeling.
Simulation results demonstrate the effectiveness of CMDP modeling and validate the safety and efficiency of the proposed pretraining framework. Specifically, compared with non-pretrained baseline, the agent pretrained through our proposed framework achieves twice the initial EE and maintains a low delay constraint violation rate of $1\%$, while ultimately converging to an EE that is $4.7\%$ higher with a $50\%$ reduction in exploration steps.
Additionally, our proposed pretraining framework implementation exhibits comparable performance to the SOTA diffusion model-based implementation, while achieving a $14$-fold reduction in computational complexity.
\end{abstract}


\section{Introduction}
\IEEEPARstart{W}{ith} 
the proliferation of mission-critical applications such as industrial automation, tele-surgery, autonomous driving, and embodied intelligence, future wireless networks are facing unprecedented demands for delay-deterministic communications, where not only ultra-low latency but also bounded jitter and high reliability must be strictly guaranteed \cite{varga2021quick, rost2022performance, zanbouri2024comprehensive, sharma2023toward}. Unlike traditional mobile broadband services, these emerging applications require guaranteed Quality of Service (QoS) in terms of per-packet delay bounds and stringent violation probabilities, pushing wireless system design beyond average-performance-centric paradigms \cite{li2022simple, schiessl2018delay}. In this context, achieving deterministic latency over inherently stochastic wireless channels has become a fundamental challenge, particularly under dynamic user demands. 

Ultra-reliable low-latency communication (URLLC) is designed to meet the stringent requirements of mission-critical applications by ensuring a block error rate below $10^{-6}$ and low latency under $1$ ms \cite{8933345}.
The current URLLC framework provides a physical layer foundation for deterministic, low-latency data transmission, utilizing short-packet transmission and finite-blocklength coding techniques \cite{10138585}.
However, despite these advancements, achieving the strict URLLC requirements remains challenging due to the limited radio resources available. To address this, agile and adaptive radio resource management strategies are essential. These include user equipment (UE) scheduling, time and frequency resource allocation, beamforming, and other technologies aimed at minimizing packet queuing and transmission delays, thereby improving the overall deterministic delay performance.

The delay guarantees in wireless systems are generally formulated in a statistical sense, typically in terms of delay bounds and violation probabilities \cite{meng2025delay}. In line with this, the Third Generation Partnership Project (3GPP) defines latency requirements using probabilistic reliability metrics \cite{3gpp38913}.
Cell-free multiple-input multiple-output (CF-MIMO) architectures have been considered to be a promising technology to enhance deterministic communication ability of wireless networks.
As a distributed wireless access architecture with cooperative access points (APs), CF-MIMO exploits macro-diversity and user-centric transmission to provide more uniform coverage. These features enhance link reliability and mitigate channel fluctuations, thereby improving service stability and offering more favorable conditions for delay-sensitive applications \cite{10145001, 10138410}.

While stringent delay guarantees are essential for mission-critical services, enforcing such requirements in wireless systems typically incurs a significant energy cost. In particular, achieving tighter delay bounds or lower violation probabilities often requires higher transmit power, increased resource allocation, and more aggressive scheduling strategies. This issue is further exacerbated in CF-MIMO systems, where cooperative transmissions among distributed APs, dynamic resource allocation, and queue-aware operations introduce additional energy consumption.
This interplay reveals a fundamental and practically relevant tradeoff between energy efficiency (EE) and delay performance. This motivates the formulation of an EE maximization problem under statistical delay constraints. 

However, solving the optimization problem in CF-MIMO systems is highly challenging. First, CF-MIMO involves high-dimensional resource allocation, including user scheduling, time-frequency resource allocation, and power control, leading to a large-scale joint optimization problem. Moreover, the system operates in a dynamic environment, where channel conditions \cite{7945856}, traffic arrivals, and queue states \cite{9541328} evolve over time, rendering the problem inherently sequential and time-coupled. Furthermore, statistical delay constraints, typically characterized by queue dynamics and delay violation probabilities, introduce additional randomness and non-convexity into the problem formulation \cite{9516890, 8964482}.

Conventional model-based optimization provides a feasible framework for wireless resource allocation \cite{10445487,ding2023joint,10045007}. However, in CF-MIMO systems with high-dimensional, dynamic, and long-term delay-constrained objectives, such approaches typically rely on simplified analytical models and incur substantial computational complexity for online implementation. As a result, it is challenging to simultaneously achieve real-time operation, adaptability to time-varying environments, and long-term performance guarantees.
To address these limitations, data-driven approaches such as deep reinforcement learning (DRL) have recently attracted significant attention, owing to their capability of handling sequential decision-making problems under complex system dynamics \cite{shi2023machine,10669755}. Nevertheless, the direct deployment of DRL in practical CF-MIMO systems faces two critical limitations: 1) the low sample efficiency of DRL, which incurs high sampling costs in real-world implementations, and 2) the cold-start exploration phase, which inevitably violates stringent delay constraints and poses significant risks to QoS-sensitive applications.


To bridge the gap between DRL's theoretical potential and practical deployment constraints, digital twin (DT) has emerged as a pivotal technology for enhancing deterministic URLLC in the $6$G era \cite{10628026}. The fundamental mechanism involves establishing a bidirectional mapping between the physical environment and its virtual counterpart, thereby facilitating efficient exploration by DRL agents through real-time synchronization and predictive simulation capabilities.



\subsection{Related Work}
Existing works related to this paper can be broadly categorized into three main directions: energy-efficiency optimization in wireless networks, statistical delay guarantee for delay-sensitive communications, and data-driven dynamic resource management enabled by DRL and digital twin technologies. These research lines are briefly reviewed as follows.

\subsubsection{\bf Energy-Efficiency Optimization in Wireless Networks}
The EE optimization has been extensively investigated in various wireless network architectures, including orthogonal frequency-division multiple access systems, MIMO networks, relay-assisted communications, and multi-cell scenarios \cite{cheung2013achieving, cheung2014spectral, jing2015flexible, cheung2015distributed, tan2017power, miao2019energy}. Prior works have explored EE-oriented designs over different control dimensions, such as power allocation, beamforming, and resource scheduling, and have commonly relied on model-based optimization tools, including fractional programming and distributed optimization, to improve energy utilization efficiency.
Nevertheless, these works predominantly focus on average performance metrics or static optimization settings, without fully accounting for statistical delay guarantees and dynamic queue evolution. When extended to CF-MIMO systems, the problem becomes significantly more challenging due to high-dimensional control and strong resource coupling. Furthermore, existing studies have yet to establish a unified framework that jointly addresses delay-aware constraints and safe learning in such complex environments.


\subsubsection{\bf Statistical Delay Guarantee for Delay-Sensitive Wireless Communications}
Statistical delay guarantees have been extensively studied for delay-sensitive wireless communications, where probabilistic QoS metrics are adopted to characterize latency performance. 
For instance, effective capacity theory has been widely adopted to characterize the maximum achievable throughput under statistical delay-QoS constraints \cite{abrao2016energy,abrao2017achieving}. In addition, QoS-constrained throughput under finite blocklength transmission has been investigated for stochastic traffic arrivals, where power control strategies are developed to maximize throughput while satisfying reliability and QoS requirements \cite{8964482}.
However, most existing studies rely on relatively structured system models and are heavily dependent on mathematical modeling and optimization techniques. When extended to CF-MIMO systems, the problem becomes significantly more complex due to high-dimensional resource coupling and stronger system dynamics. In such scenarios, the resulting optimization problems often suffer from high computational complexity and limited scalability, making real-time implementation and adaptation to dynamic environments particularly challenging.

\subsubsection{\bf DRL and Digital-Twin-Enabled Dynamic Resource Management}
Recent advances have formulated statistical-QoS-constrained radio resource allocation as a constrained Markov decision process (CMDP), where DRL is employed to handle long-term sequential optimization, and constraints are typically managed through Lagrangian relaxation \cite{zhang2022user} or Lyapunov-based techniques \cite{wu2020accuracy}. This provides a flexible framework for dynamic wireless resource management under time-varying channels, traffic arrivals, and queue evolution.


Meanwhile, current DT implementations predominantly follow two distinct paradigms. The first approach employs parametric modeling with continuous state synchronization \cite{10371218,9982429,10319784}, requiring periodic updates for environment-twin consistency. While this method enhances resource management through real-time state awareness, its scalability in large-scale networks becomes constrained by escalating communication overhead. As demonstrated in \cite{9982429}, exponential data transmission demands create substantial communication loads, making delay-sensitive parameter updates increasingly challenging.
The second paradigm establishes environment mappings through neural networks (NNs) trained on historical operational data \cite{10522623,10225588,10132044,xiong2025enhancing,chai2024generative}. This methodology supports DRL exploration through high-fidelity synthetic data generation, though it faces challenges in addressing distribution shifts between historical training data and current environmental states. A notable implementation in \cite{10522623} combines long short-term memory networks for state transition prediction with deep NNs for reward estimation, creating a virtual pre-verification environment for DRL exploration in network slicing scenarios.

The remarkable success of generative NNs across diverse domains, e.g., image synthesis \cite{ho2020denoising}, natural language processing \cite{devlin2019bert}, and symbolic music generation \cite{9740506}, has inspired novel DT architectures.
Unlike conventional deterministic mapping NNs, generative NNs demonstrate superior multimodal generation capability through the learning architecture for probability distribution.  
Recent studies showcase various generative NN-powered DTs implementations.
\cite{10225588} employs variational auto-encoder (VAE) and generative adversarial network (GAN) to synthesize unselected channel states in non-terrestrial networks, accelerating DRL convergence through enhanced state exploration.
Authors in \cite{10132044} develop a cycle-GAN-based DT scheduling model for time-sensitive applications, enabling multi-stream forwarding delay prediction. Furthermore, diffusion model-powered DTs for 6G networks have been proposed in \cite{xiong2025enhancing, chai2024generative}, both of which demonstrate high fidelity and effective policy enhancement capabilities.

Despite these advances, most existing DT-enhanced DRL studies primarily focus on task-specific communication indicator mapping, while insufficient attention has been devoted to foundational DRL interaction frameworks. Specifically, the structural mapping of CMDP components that govern environment-agent dynamics requires further investigation to better accommodate the scalability and heterogeneity demands of 6G CF-MIMO scenarios.

\subsection{Motivation and Contributions}
The above review reveals several limitations in the existing literature. First, current studies on EE optimization have not adequately incorporated statistical delay guarantees, especially in wireless systems with dynamic traffic and queue evolution. Second, existing works on statistical delay guarantees are mostly developed for relatively structured scenarios and are rarely tailored to the high-dimensional and dynamic resource control problem in CF-MIMO systems. Third, although DRL- and DT-based approaches provide a promising means of handling dynamic optimization, they remain insufficient in terms of safe pretraining, sample efficiency, and structured modeling of key CMDP components. 

Motivated by these limitations, in this paper, we investigate EE optimization in CF-MIMO systems under probabilistic delay guarantee constraints and formulate the problem as a CMDP. To alleviate the low sample efficiency of online DRL and reduce the risk of delay-constraint violations during exploration, we develop a virtual CMDP for pretraining. The proposed pretraining framework enables effective offline training, thereby providing a well-initialized policy and significantly accelerating convergence during online deployment.
In particular, an evidence-aware conditional Gaussian Mixture Model (EA-CGMM) algorithm is proposed to tackle data sparsity and distribution shift in modeling state transition probabilities of the virtual CMDP.
The main contributions of this paper are as follows:

\begin{itemize}
	\item 
	We address the space-time-frequency resource allocation problem in a CF-MIMO downlink transmission system, aiming to maximize EE subject to a constraint on the delay violation rate. We formulate the optimization problem as a CMDP and solve it using Proximal Policy Optimization (PPO) with a primal-dual method.
	\item 
	We propose a novel offline pretraining framework based on the robust CMDP modeling. It collects offline transition tuples from the environment CMDP to drive the reward and cost function module, the initial-state distribution module, and the state transition module to form a virtual CMDP, enabling controllable policy optimization while improving the sample efficiency.
	\item 
	To overcome data sparsity and distribution shift in state transition modeling,
	we propose the EA-CGMM inference approach.
	The approach effectively infers the next-state distribution through analytical derivation for the explicit GMM modeling of state transition tuple, with statistical evidence reweighting from the marginal probability of current state-action pair. 
	\item 
	The simulation results show the efficiency of the proposed virtual CMDP module implementations. In particular, when compared with the non-pretrained baseline, the pretrained agent enters the real environment with twice the initial EE and a delay constraint violation rate as low as $1\%$, while ultimately converging to an EE that is $4.7\%$ higher and requiring $50\%$ fewer exploration steps.
\end{itemize}

It is worth noting that the proposed virtual CMDP modeling and pretraining framework are not tied to specific structural assumptions of CF-MIMO systems and therefore exhibits a certain degree of generality. In this work, the CF-MIMO scenario is adopted as a representative and challenging large-scale network architecture to demonstrate the effectiveness of the proposed method. The framework can be readily extended to other wireless communication settings, such as conventional co-located massive MIMO systems or multi-cell networks.

The rest of this paper is organized as follows. The system model is introduced in Section \ref{SystemModel}. In Section \ref{ProblemFormulation}, the resource management problem is formulated as a CMDP. In Section \ref{PretrainingFramework}, the pretraining framework is detailed. The simulation results and analysis are presented in Section \ref{Expiriment} and this paper is summarized in Section \ref{Conclusion}.
\textit{Notation:}
The superscripts $\left(\cdot\right)^T$ and $\left(\cdot\right)^\mathrm{H}$ denote the transpose and conjugate transpose operations, respectively. $\mathcal{CN}\left(\boldsymbol{a},\boldsymbol{B}\right)$ denotes a complex Gaussian distribution with mean
vector $\boldsymbol{a}$ and covariance matrix $\boldsymbol{B}$.
$\left|x\right|$ and $\left|\boldsymbol{X}\right|$ represent the magnitude of the scalar $x$ and determinant of the matrix $\boldsymbol{X}$.
The $l_1$-norm and $l_2$-norm are expressed by $\left\|\cdot\right\|_1$ and $\left\|\cdot\right\|_2$.
The expectation operation is denoted by $\mathbb{E}\left[\cdot\right]$. 
The function $\text{diag}(\cdot)$ extracts the diagonal elements of a matrix to form a vector, and $\mathbb{I}\left[\cdot\right]$ denotes the indicator function, which returns $1$ if the condition inside the brackets is true, and $0$ otherwise.
Additionally, $\mathcal{U}[a,b]$ denotes the continuous uniform distribution over $[a,b]$.
For ease of reference, the key notations are listed in Table \ref{key notations}.
\begin{table}
	\caption{Key Notations}
	\label{key notations}
	\centering
	\begin{tabularx}{\linewidth}{l X}
		\hline
		\textbf{Notations} & $\textbf{Meanings}$ \\
		\hline
		$s_u^{t,k,n,c}$ &  Modulated signal with unit energy for UE $u$ in RE $\{t, k, n, c\}$ \\
		$\zeta_{b,u}^{t,k}$    & Binary scheduling indicator of UE $u$ in TFU $\{t,k\}$ of AP $b$   \\
		$P_{b,u}^{t,k}$  & Power allocated by AP $b$ to UE $u$ in TFU $\{t,k\}$ (W)  \\
		$\mathbf{f}_{b,u}^{t,k}$  & Beamforming vector of AP $b$ for UE $u$ in TFU $\{t,k\}$ \\
		$\mathbf{h}_{b,u}^{t,k}$  & Downlink channel of AP $b$ to UE $u$ in TFU $\{t,k\}$ \\
		$\gamma_u^{t, k}$ & SINR of UE $u$ over all REs within TFU $\{t, k\}$\\
		$\psi_u^t$ & Number of bits transmitted to UE $u$ in time slot $t$\\
		$A_{u}(t)$ & Number of packets arriving at time slot $t$ for UE $u$ \\
		$G_{u}^{a,t}$ & Size of the $a$-th arriving packet for UE $u$ at slot $t$ (in bits) \\
		$Q_{u}^{\max}$ & Maximum buffer capacity of UE $u$ (in bits)\\
		$Q_{u}(t)$ & Queue length at the beginning of slot $t$ (in bits)\\
		$d_{u,a}(t)$ & Delay for the $a$-th packet arriving at slot $t$ (in time slots)\\
		$D_u$ & Deadline of UE $u$ (in time slots) \\
		$\eta_u$ & Threshold of packet delay violation probability for UE $u$ \\
		$\rho$ & Initial state distribution of the CMDP\\
		$\mathcal{T}$ & State transition function of the CMDP\\
		$r$ & Reward function of the CMDP\\ 
		$c$ & Cost function of the CMDP\\
		$\boldsymbol{r}_t$ & PBM vector representing beam domain channel state\\
		$\boldsymbol{q}_t$ & Queue state vector including buffer occupancy and urgency information\\ 
		$\pi_g$ & Mixture weights of the $g$-th mixture component in the GMM\\
		$\boldsymbol{\mu}_g$ & Mean vector of the $g$-th mixture component in the GMM\\
		$\boldsymbol{U}_g$ & Cholesky factor of the $g$-th mixture component in the GMM \\
		$\boldsymbol{\Sigma}_g$ & Covariance matrix of the $g$-th mixture component in the GMM \\
		$\chi^2_{d, 1-\alpha}$ & Upper $(1-\alpha)$-quantile of the chi-squared distribution with $d$ degrees of freedom\\
		$\boldsymbol{m}$ & Statistical significance mask vector\\ 
		\hline
	\end{tabularx}
\end{table}

\section{System Model} \label{SystemModel}
We consider a CF-MIMO orthogonal frequency division multiplexing (OFDM) downlink system, where $B$ APs each with $M$ antennas cooperatively serve $U$ single-antenna users. The system frequence-band with bandwidth $B_w$ Hz is divided into $K$ subbands. Each subband is the smallest schedulable frequency unit and consists of $C$ consecutive subcarriers. The subcarrier interval is then $\frac{B_w}{KC}$ (Hz) and the OFDM symbol time is $\frac{KC}{B_w}$ (s). Define the smallest schedulable  time unit as one slot which consists of $N$ OFDM symbols, i.e., the slot length is $\triangle t_s= N\frac{KC}{B_w}$ (s). Therefore, one minimum schedulable time-frequency unit (TFU) (i.e, one subband in one slot) has $CN$ resource elements (REs). Denote the sets of APs, users and subbands as $\mathbb{B} = \left\{ 1, \ldots, B \right\}$, $\mathbb{U} = \left\{ 1, \ldots, U \right\}$ and $\mathbb{K} = \left\{ 1, \ldots, K \right\}$ respectively.

\subsection{Transmission Model}
In the downlink signal transmission stage, APs collaboratively transfer the signals to the scheduled users over the assigned REs. We consider the transmission within a time window consisting of $T$ time slots, where the set of time slots is denoted as $\mathbb{T} = \{ 1, \ldots, T \}$. The binary variable $\zeta_{b, u}^{t,k}$ indicates whether UE $u$ is scheduled in TFU $\{t, k\}$ of AP $b$, where $b \in \mathbb{B}, u \in \mathbb{U}, k \in \mathbb{K}, t \in \mathbb{T}$.

Assume that the channel variation within one TFU can be ignored. In practice, the optimization granularity for power and the beamforming can be no smaller than the TFU level. Specifically, the power is equally divided among all REs belonging to the same TFU and the beamforming settings for all REs belonging to the same TFU are also the same.

Each RE $\{n, c\}$ within TFU $\{t, k\}$ is denoted as RE $\{t, k, n, c\}$, where $n \in \mathbb{N}$ and $c \in \mathbb{C}$ represent the OFDM symbol and subcarrier indices, respectively. Given the beamforming vector of AP $b$ for UE $u$, i.e., $\mathbf{f}_{b, u}^{t, k} \in \mathbb{F}$,  where $\mathbb{F} = \{ \mathbf{f}_{1}, \ldots, \mathbf{f}_{M}\} \in \mathbb{C}^{M \times M}$ is the beamforming codebook with $M$ codewords satisfying $\|\mathbf{f}_m\|= 1, m \in \mathbb{M} = \{1, \ldots, M \}$, and the power allocated by AP $b$ to UE $u$, i.e., $\frac{P_{b, u}^{t, k}}{C}$, the transmitted signal from AP $b$ can be expressed as
\begin{equation}
	\mathbf{x}_b^{t,n,k,c}=\sum_{u\in\mathbb{U}}\zeta_{b,u}^{t,k}\sqrt{\frac{P_{b,u}^{t,k}}{C}}\mathbf{f}_{b,u}^{t,k}s_u^{t,k,n,c},
\end{equation}
where $s_u^{t, n, k, c}$ is the modulated symbol with unit energy for UE $u$, i.e., $\mathbb{E}\{\left|s_u^{t,n,k,c}\right|^2\} = 1$.
In RE $\{t, n, k, c \}$, the received signal of UE $u$ is modeled as in Eq. \eqref{received_signal}, where $\mathbf{h}_{b,u}^{t,k} \in \mathbb{C}^{M \times 1}$ represents the downlink channel and $e_u^{t,k,n,c} \sim \mathcal{CN}(0, \sigma_u^2)$ denotes the additive white Gaussian noise (AWGN) at UE $u$, with $\sigma_u^2$ being the noise variance.
\begin{figure*}[ht] 
\centering 
\begin{equation}
	\label{received_signal}
	\begin{aligned}
		y_{u}^{t,k,n,c}& =\sum_{b\in\mathbb{B}}\sum_{v\in\mathbb{U}}\zeta_{b,v}^{t,k}\sqrt{\frac{P_{b,u}^{t,k}}{C}}\mathbf{h}_{b,u}^{t,k,\mathrm{H}}\mathbf{f}_{b,v}^{t,k}s_v^{t,k,n,c}+e_u^{t,k,n,c} \\
		&=\underbrace{\sum_{b\in\mathbb{B}}\zeta_{b,u}^{t,k}\sqrt{\frac{P_{b,u}^{t,k}}{C}}\mathbf{h}_{b,u}^{t,k,\mathrm{H}}\mathbf{f}_{b,u}^{t,k}s_{u}^{t,k,n,c}}_{{\text{desired signal}}}+\underbrace{\sum_{b\in\mathbb{B}}\sum_{{v\in\mathbb{U},v\neq u}}\zeta_{b,v}^{t,k}\sqrt{\frac{P_{b,v}^{t,k}}{C}}\mathbf{h}_{b,u}^{t,k,\mathrm{H}}\mathbf{f}_{b,v}^{t,k}s_{v}^{t,k,n,c}}_{{\text{interference}}}+e_{u}^{t,k,n,c}.
\end{aligned}\end{equation}
\end{figure*}
The signal-to-interference-plus-noise ratio (SINR) of UE $u$ over all REs within TFU $\{t, k\}$ remains the same, given by
\begin{equation}
	\label{SINR}
	\gamma_u^{t, k}=\frac{\left\vert\sum_{b\in\mathbb{B}}\zeta_{b,u}^{t,k}\sqrt{\frac{P_{b,u}^{t,k}}{C}}\mathbf{h}_{b,u}^{t,k,\mathrm{H}}\mathbf{f}_{b,u}^{t,k}\right\vert^2}{\sum_{v\in\mathbb{U},v\neq u}\left\vert\sum_{b\in\mathbb{B}}\zeta_{b,v}^{t,k}\sqrt{\frac{P_{b,v}^{t,k}}{C}}\mathbf{h}_{b,u}^{t,k,\mathrm{H}}\mathbf{f}_{b,v}^{t,k}\right\vert^2+\sigma_u^2}.
\end{equation}

Considering the practical finite blocklength coding, given the block error probability $\epsilon_u$, the number of bits transmitted to UE $u$ in time slot $t$ is\cite{meng2025delay}
\begin{equation}
	\psi_u^t=\sum_{k\in\mathbb{K}}CN\log_2\left(1+\gamma_u^{t,k}\right)-Q^{-1}\left(\epsilon_u\right)\sqrt{\sum_{k\in\mathbb{K}}CNV_u^{t,k}},
\end{equation}
where $Q^{-1}\left(\right)$ is the inverse of the Gaussian Q-function, and $V_u^{t,k}$ represents the channel dispersion, calculated as
\begin{equation}V_u^{t,k}=\left[\log_2\left(e\right)\right]^2\left(1-\left(1+\gamma_u^{t,k}\right)^{-2}\right).\end{equation}

\subsection{Queueing and Delay Model}
Let $A_{u}(t)$ denote the number of packets arriving at time slot $t$ for UE $u$, which is a random variable. The $a$-th arriving packet size (in bits) for UE $u$ at slot $t$ is denoted as $G_{u}^{a,t}$, where $a\in\{1,2,\ldots,A_{u}(t)\}$. A general assumption is that $A_{u}(t)$ is independently and identically distributed (i.i.d.) over times slots with mean arrival rate $\lambda_u$. We consider that all packets arriving at the same UE $u$ share a common deadline, measured in time slots and denoted by $D_u$.
Each UE $u$ maintains a finite buffer with maximum capacity $Q_{u}^{\max}$ bits. Let $Q_{u}(t)$ represent the queue length (bits) at the beginning of slot $t$, evolving according to:
\begin{equation}
	Q_{u}(t+1)=\min\left\{[Q_{u}(t)-\psi_u^t]^{+} + \sum\nolimits_{a=1}^{A_{u}(t)}G_u^{a, t},Q_{u}^{\max}\right\},
\end{equation}
where $[x]^{+}\triangleq\max\{x, 0\}$. Packets arriving when $Q_u(t) \geq Q_u^{\max}$ are immediately dropped. Existing packets are scheduled via an earliest deadline first (EDF) policy with first-in-first-out (FIFO) discipline within each UE's buffer.
For the $a$-th packet arriving at slot $t$ (if not dropped), its delay (in time slots) is
\begin{equation}
	\begin{aligned}
		d_{u,a}(t) = \underset{\tau}{\operatorname*{\arg\min}} \biggl\{ \left[Q_{u}(t) - \psi_{u}^{t}\right]^{+} + \sum\nolimits_{i=1}^{a}G_{u}^{i,t} \biggr. 
		\\ 
		\phantom{=\;\;}
		\biggl. - \sum\nolimits_{i=t+1}^{\tau} \psi_u^i \leq 0 \biggr\} - t.
	\end{aligned}
\end{equation}
Dropped packets are excluded from delay calculations.

\section{Problem Formulation and Solution}\label{ProblemFormulation}
The objective is to find a joint space-time-frequency resource allocation policy for TFU resources, beamforming vectors, and power allocation that maximizes the EE of the CF MIMO-OFDM downlink system described in Section \ref{SystemModel}, while satisfying a delay violation rate constraint.
The problem can be formulated as follows:
\begin{subequations}
	\begin{align}
		\max_{\{\zeta_{b,u}^{t,k},\mathbf{f}_{b,u}^{t,k},P_{b,u}^{t,k}\}} & f_{\mathrm{EE}}\left(\left\{\zeta_{b,u}^{t,k},\mathbf{f}_{b,u}^{t,k},P_{b,u}^{t,k}\right\}\right) \label{eq:objective} \\
		\mathrm{s.t.} & \Pr\left(d_{u,a}\left(t\right)>D_u\right)<\eta_u, \quad \forall u\in\mathbb{U}, \forall t\in\mathbb{T},  \label{eq:constraint1} \\
		& \sum_{k\in\mathbb{K}}\sum_{u\in\mathbb{U}}\zeta_{b,u}^{t,k}P_{b,u}^{t,k}\leq P_{\max}, \quad \forall b\in\mathbb{B}, \forall t\in\mathbb{T}, \label{eq:constraint2} \\
		& \zeta_u^{t,k}\in\left\{0,1\right\}, \quad \forall u\in\mathbb{U}, \forall t\in\mathbb{T}, \forall k\in\mathbb{K}, \label{eq:constraint3} \\
		& \mathbf{f}_{b,u}^{t,k}\in\mathbb{F}, \quad \forall u\in\mathbb{U}, \forall t\in\mathbb{T}, \forall k\in\mathbb{K}, \label{eq:constraint4}
	\end{align}
\end{subequations}
where the EE in Eq. \eqref{eq:objective} is calculated as:
\begin{equation}\label{opt_EE}
	f_{\mathrm{EE}}\biggl(\!\Bigl\{\zeta_{b,u}^{t,k},\mathbf{f}_{b,u}^{t,k},P_{b,u}^{t,k}\Bigr\}\!\biggr) 
	= \frac{
		\sum\limits_{t\in\mathbb{T}} \sum\limits_{u\in\mathbb{U}} \psi_u^t
	}{
		\sum\limits_{t\in\mathbb{T}} 
		\sum\limits_{b\in\mathbb{B}} 
		\sum\limits_{k\in\mathbb{K}} 
		\sum\limits_{u\in\mathbb{U}} 
		\zeta_{b,u}^{t,k} P_{b,u}^{t,k} \triangle t_s
	}.
\end{equation}
Constraint \eqref{eq:constraint1} ensures that the packet delay violation probability of each UE remains below its corresponding threshold $\eta_u$
\footnote{This probabilistic reliability constraint is consistent with the reliability–latency tradeoff paradigm in 3GPP URLLC \cite{3gpp38913} and with the statistical delay guarantee formulation commonly adopted in wireless systems \cite{meng2025delay}. 
}.
Constraint \eqref{eq:constraint2} enforces per-AP power budgets $P_{\max}$ across all subcarriers and UEs. Binary variable constraint \eqref{eq:constraint3} governs discrete TFU resource allocation, while \eqref{eq:constraint4} restricts beamforming vectors to predefined codewords from the available codebook $\mathbb{F}$.


\subsection{Constrained Markov Decision Process}
The resource management problem can be formulated as a CMDP, denoted by $\mathcal{M} \cup \mathcal{C}$. The standard MDP is represented by $\mathcal{M}=\langle\mathcal{S},\mathcal{A},\mathcal{T},\rho, \gamma,r\rangle$, where $\mathcal{S}$ and $\mathcal{A}$ denote the state space and action space, respectively. The state transition function is denoted by $\mathcal{T}\left(s,a\right)$. $\rho$ represents the initial-state distribution. The discount factor is denoted by $\gamma$ and reward function is given by $r(\boldsymbol{s},\boldsymbol{a}): \mathcal{S}\times\mathcal{A}\to\mathbb{R}$. 
The constraint term is incorporated through $\mathcal{C}=\langle c, d \rangle$, where $c(\boldsymbol{s},\boldsymbol{a}): \mathcal{S}\times\mathcal{A}\to\mathbb{R}$ represents cost function and $d\in\mathbb{R}$ specifies cost threshold. While a single constraint is considered here for simplicity, the formulation can be readily extended to multiple constraints.

The objective is to learn an optimal policy $\pi^{*}$ that maximizes the expected return while satisfying the safety constraints. This can be mathematically expressed as:
\begin{equation}
	\pi^*=\arg\max_{\pi\in\Pi_{c}}V_r^\pi(\rho),
\end{equation}
where $V_r^\pi(\rho)$ is expected discounted return of a fixed policy $\pi$ with respect to the initial-state distribution $\rho$, defined as:
\begin{equation}\label{ValueFunctionR}
	V_r^\pi(\rho) :=
	\mathbb{E}_{\substack{
			\boldsymbol{s}_0 \sim \rho \\
			\boldsymbol{a}_t \sim \pi(\cdot | \boldsymbol{s}_t) \\
			\boldsymbol{s}_{t+1} \sim \mathcal{T}(\cdot | \boldsymbol{s}_t, \boldsymbol{a}_t)
	}}
	\left[
	\sum\nolimits_{t=0}^\infty \gamma^t r\left(\boldsymbol{s}_t, \boldsymbol{a}_t\right)
	\right].
\end{equation}
The safety-constrained policy set is formally defined as the collection of all policies satisfying the safety constraints:
\begin{equation}\label{ValueFunctionC}
	\Pi_{c}=\{\pi\in\Pi: V_{c}^\pi(\rho)<d\},
\end{equation}
where $\Pi$ denotes the set of all feasible policies, and $V_{c}^\pi(\rho)$ is expected discounted cost, analogous to $V_r^\pi(\rho)$ but with reward function $r\left(\mathbf{s}_t,\mathbf{a}_t\right)$ replaced by the cost function $c\left(\mathbf{s}_t,\mathbf{a}_t\right)$.

\subsection{Problem-Specific Design} \label{problem-specific}
\subsubsection{State Space}
In time slot $t$, the state $\mathbf{s}_t$ comprises both channel and queue status information. 
The channel state is represented by the probing beam measurement (PBM) vector $\boldsymbol{r}_t \in \mathbb{R}^{B \cdot U \cdot K \cdot M}$, obtained by vectorizing the set $\left\{r_{b,u}^{t,k,m} \mid b \in \mathbb{B}, u \in \mathbb{U}, k \in \mathbb{K}, m \in \mathbb{M}\right\}$, where $r_{b,u}^{t,k,m} = |\mathbf{h}_{b,u}^{t,k,\mathrm{H}} \times \mathbf{f}_{b,u}^{t,k,m}|$ quantifies the beamforming gain for each subband between AP-UE pairs.
The queue state is given by $\boldsymbol{q}_t=[\boldsymbol{q}_t^{\text{buf}}, \boldsymbol{q}_t^{\text{urg}}] \in \mathbb{R}^{2U}$, where
$\boldsymbol{q}_t^{\text{buf}}=[q_{t,1}^{\text{buf}}, q_{t,2}^{\text{buf}}, \ldots, q_{t,U}^{\text{buf}}]$ represents buffer occupancy in terms of packet counts and $\boldsymbol{q}_t^{\text{urg}}=[q_{t,1}^{\text{urg}}, q_{t,2}^{\text{urg}}, \ldots, q_{t,U}^{\text{urg}}]$ captures latency-critical traffic through the number of packets with delay reaching the deadline $D_u$.
This composite state representation $\boldsymbol{s}_t=(\boldsymbol{r}_t,\boldsymbol{q}_t)$
provides a comprehensive characterization of current network conditions for the data-driven decision-making. 


\subsubsection{Action Space}
The action space $\mathbb{A}= \left\{ \hat{\zeta}_{b,u}^{t,k}, \hat{i}_{b,u}^{t,k}, \hat{P}_{b,u}^{t,k} \mid b \in \mathbb{B}, u \in \mathbb{U}, t \in \mathbb{T}, k \in \mathbb{K} \right\}$ is generated by the agent's policy network through three continuous outputs that undergo transformations to produce discrete TFU allocations and beamform selections, as well as practical power allocations. Specifically, thresholds are applied to transform $\hat{\zeta}_{b,u}^{t,k}\in(-1, 1)$ and $\hat{i}_{b,u}^{t,k}\in(-1, 1)$ into the TFU allocation variable $\zeta_{b,u}^{t,k} \in \left\{0, 1\right\}$ and the beamforming vector index $i_{b,u}^{t,k} \in \mathbb{M}$, mitigating the combinatorial complexity. The floor function is defined as $\left\lfloor x \right\rfloor$, which returns the largest integer less than or equal to $x$. Consequently, we have:
\begin{equation}
	\zeta_{b,u}^{t,k} = \left\lfloor \frac{\hat{\zeta}_{b,u}^{t,k} + 1}{2} \right\rfloor,
\end{equation}
\begin{equation}
	i_{b,u}^{t,k} = \left\lfloor \frac{\left(\hat{i}_{b,u}^{t,k} + 1\right) \cdot (M - 1)}{2} \right\rfloor.
\end{equation}
Let $\hat{\boldsymbol{P}}_{b}^{t}=\{\hat{P}_{b,u}^{t,k} \mid u \in \mathbb{U}, k \in \mathbb{K}\}$ denote the power allocation ratio vector for AP $b$, where $\hat{P}_{b,u}^{t,k} \in \left(-1, 1\right)$.
To strictly satisfy each AP's maximum power constraint, the actual power allocation $\boldsymbol{P}_{b}^{t}$ can be obtained through
\begin{equation}
	\boldsymbol{P}_{b}^{t}=P_{\max}\cdot\mathrm{Softmax}(\hat{\boldsymbol{P}}_{b}^{t}), \quad \forall b\in\mathbb{B}, \forall t\in\mathbb{T}.
\end{equation}

\subsubsection{State Transition Function}
The state transition function $\mathcal{T}\left(\boldsymbol{s}_{t+1}\mid \boldsymbol{a}_t,\boldsymbol{s}_t\right)$ specifies probability distribution of transitioning to next-state  $\boldsymbol{s}_{t+1}$ given current state $\boldsymbol{s}_t$ and action $\boldsymbol{a}_t$. As the equivalent channel under the complete beamforming codebook $\mathbb{F}$ is considered, there is no coupling between the channel state and beam selection. Exploiting the independence between the channel state and the queue state, the transition function can be decomposed into the channel state transition $\mathcal{T}_{\boldsymbol{r}}\left(\boldsymbol{r}_{t+1}\mid \boldsymbol{r}_t\right)$, which is independent of the action, and the queue state transition $\mathcal{T}_{\boldsymbol{q}}\left(\boldsymbol{q}_{t+1}\mid \mathbf{a}_t,\boldsymbol{s}_{t}\right)$.

\subsubsection{Reward Function}
In order to achieve the optimization goal of maximizing the average EE of Eq. \eqref{opt_EE}, the reward function is set as the single-slot EE, i.e.,
\begin{equation}\label{r_EE}
	r\left(\boldsymbol{s}_t,\boldsymbol{a}_t\right)=\frac{\sum_{u\in\mathbb{U}}\psi_u^t}{\sum_{b\in\mathbb{B}}\sum_{k\in\mathbb{K}}\sum_{u\in\mathbb{U}}\zeta_{b,u}^{t,k}P_{b,u}^{t,k}\triangle t_s}.
\end{equation}

\subsubsection{Constraint Term}

According to Constraint \eqref{eq:constraint1}, for each user $u$, cost threshold is set to $d_u=\eta_u$ and cost function is defined as the single-slot delay constraint violation rate, i.e.,
\begin{equation}
	c_u(\boldsymbol{s}_t,\boldsymbol{a}_t) =
	\frac{q_{t,u}^{\text{vio}} + q_{t,u}^{\text{drop}}}
	{q_{t,u}^{\text{vio}} + q_{t,u}^{\text{drop}} + q_{t,u}^{\text{tx}}},
	\quad \forall u\in\mathbb{U},
\end{equation}
where $q_{t,u}^{\text{vio}}$, $q_{t,u}^{\text{drop}}$ and $q_{t,u}^{\text{tx}}$ represent the numbers of delay-violated packets, dropped packets due to capacity exceedance and transmitted packets, respectively, in time slot $t$.
In this work, per-user constraint term $(c_u-d_u)$ is aggregated into a single system-level average.\footnote{This aggregation enables the use of a single Lagrange multiplier (see Section~\ref{larg}) for constraint handling, leading to faster convergence and better training stability, but relaxes the delay violation probability requirements for individual users.}
Furthermore, constraints \eqref{eq:constraint2}, \eqref{eq:constraint3}, and \eqref{eq:constraint4} are inherently satisfied through the design of the action space, thus eliminating the need for additional handling in the optimization process.

\subsection{Primal-Dual Approach}\label{larg}
We adopt the typical primal-dual approach to solve the CMDP problem, which transforms the constrained problem into an unconstrained Lagrangian formulation by introducing a weighted constraint term into the objective function. The Lagrangian function is defined as:
\begin{equation}
	\min_{\lambda\in\mathbb{R}^+}\max_{\pi\in\Pi}\mathcal{L}(\pi,\lambda):=V_r^\pi(\rho)-\lambda\left(V_{c}^\pi(\rho)-d\right),
\end{equation}
where $\lambda \in \mathbb{R}^+$ is the dual variable.

The optimization process involves alternating between policy gradient updates and gradient descent updates of the dual variable $\lambda$. Specifically, at training step $t$, the policy $\pi_{t+1}$ is first updated according to: 
\begin{equation}
\pi_{t+1}\leftarrow\pi_t+\alpha_\pi\left(\nabla_\pi V_r^{\pi_t}(\rho)-\lambda_{t}\nabla_\pi V_{c}^{\pi_{t}}(\rho)\right).
\end{equation}
Subsequently, the dual variable $\lambda_t$ is updated through:
\begin{equation}
	\lambda_{t+1}\leftarrow\max\left(\lambda_{t}+\alpha_{\lambda}\left( V_{c}^{\pi_{t}}(\rho)-d\right),0\right),
\end{equation}
where $\alpha_{\pi}$ and $\alpha_{\lambda}$ are the learning rates for policy and dual variable, respectively.

Denote the policy parameters as $\theta$. To efficiently optimize the policy $\pi_\theta$ in the continuous, high-dimensional action space, we adopt the policy gradient algorithm PPO \cite{schulman2017proximalpolicyoptimizationalgorithms} due to its stability and sample efficiency. PPO replaces conventional policy gradient updates with a clipped surrogate objective:
\begin{equation}
	\mathcal{L}^{\text{CLIP}}(\pi_\theta) = \mathbb{E}_t \left[ \min\left(\eta_t(\theta) \hat{A}_t, \text{clip}(\eta_t(\theta), 1-\varsigma, 1+\varsigma) \hat{A}_t \right) \right],
\end{equation}
where $\eta_t(\theta)=\frac{\pi_{\theta^\prime}(a_t|s_t)}{\pi_{\theta}(a_t|s_t)}$ is probability ratio between the new and old policies, $\hat{A}_t$ is the estimated advantage function, and $\varsigma$ is a hyperparameter that controls the clipping range. By constraining policy updates within a trust region to limit deviation from the previous policy, PPO mitigates the instability issues commonly encountered in traditional policy gradient methods.

\section{An Offline Pretraining Framework Based on Robust CMDP Modeling}
\label{PretrainingFramework}
Recall that the resource management problem is formulated as a CMDP, denoted as $\mathcal{M} \cup \mathcal{C} = \langle\mathcal{S},\mathcal{A},\mathcal{T}, \rho, \gamma,r,c,d\rangle$. 
As established in Eq. \eqref{ValueFunctionR} and Eq. \eqref{ValueFunctionC}, the performance of a policy $\pi$ in this CMDP, measured by its expected cumulative reward $V_r^\pi(\rho)$ and constraint satisfaction $V_{c}^\pi(\rho)<d$, is determined by the environmental parameters $(\mathcal{T}, r, \rho, c, d)$.
This dependency motivates the construction of a virtual CMDP that maps $(\mathcal{T}, r, \rho, c, d)$ with high precision. By pretraining policies in this virtual twin, we can mitigate safety risks arising from suboptimal exploration and constraint violations during real-world deployment while improving the sample efficiency.

\begin{figure}
	\centering
	\includegraphics[width=3.3in]{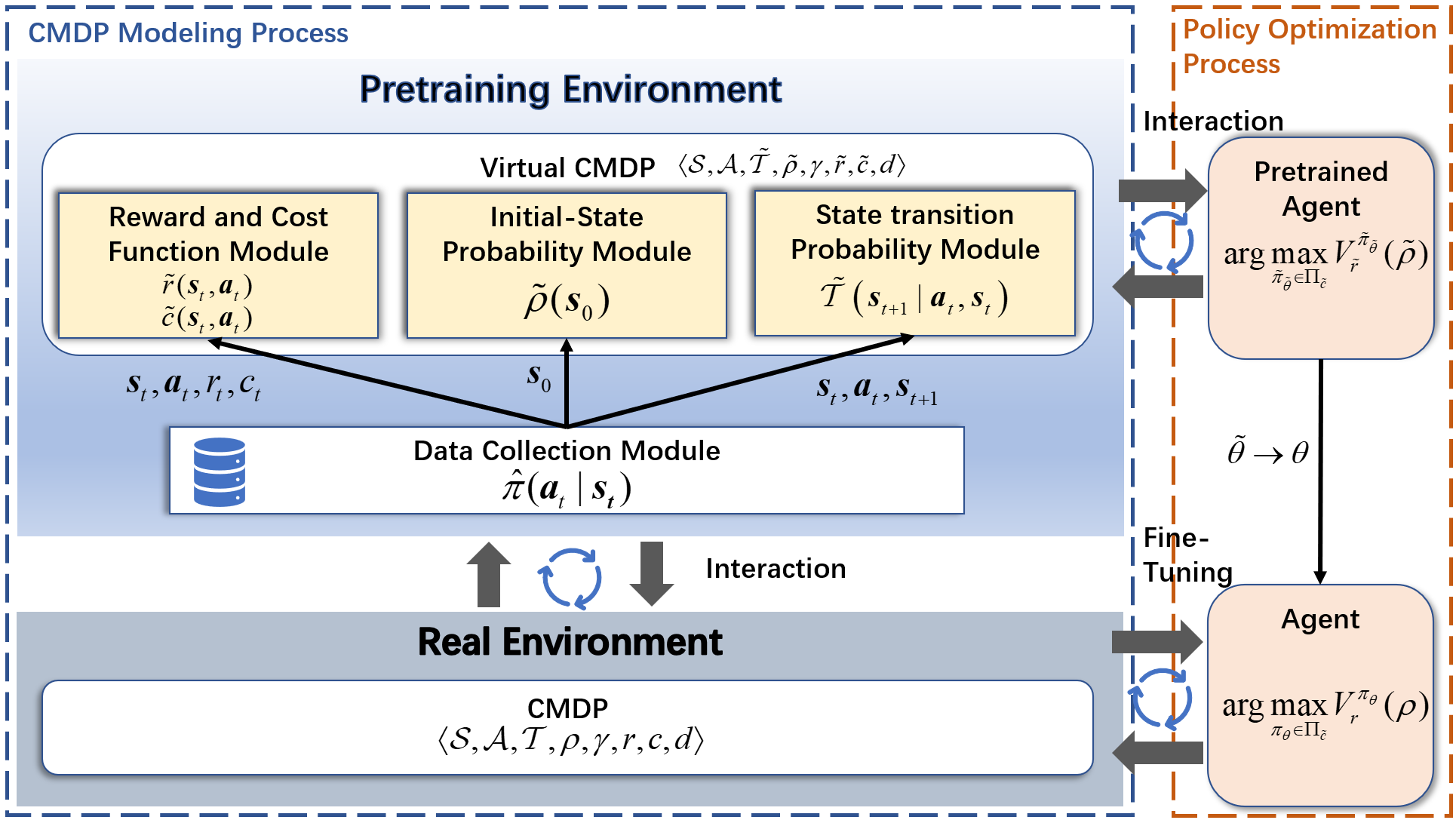}
	\caption{Proposed pretraining framework based on CMDP modeling.}
	\label{framework}
\end{figure}

As illustrated in Fig. \ref{framework}, the proposed pretraining framework operates through two interconnected phases.
The first is the CMDP modeling phase. The data collection module interacts with the real environment under an offline behavior policy $\hat{\pi}$, gathering transition tuples $\left(\boldsymbol{s}_t, \boldsymbol{a}_t, r_t, c_t, \boldsymbol{s}_{t+1}\right)$. 
These tuples are utilized to train three neural network modules:
\begin{itemize}
	\item 
	\textit{Reward and Cost Function Module}: Learns $\tilde{r}(\boldsymbol{s}_t, \boldsymbol{a}_t)$ and $\tilde{c}(\boldsymbol{s}_t, \boldsymbol{a}_t)$.
	\item 
	\textit{Initial-State Probability Module}: Estimates $\tilde{\rho}(\boldsymbol{s}_0)$.
	\item 
	\textit{State Transition Probability Module}: Predicts $\tilde{\mathcal{T}}\left(\boldsymbol{s}_{t+1}\mid\boldsymbol{a}_t, \boldsymbol{s}_t\right)$.
\end{itemize}
These modules collectively form a virtual CMDP $\tilde{\mathcal{M}} \cup \tilde{\mathcal{C}} = \langle\mathcal{S},\mathcal{A},\tilde{\mathcal{T}}, \tilde{\rho}, \gamma,\tilde{r},\tilde{c},d\rangle$, establishing a closed-loop pretraining environment that mirrors real-world dynamics.

The second is the policy optimization phase. The DRL agents interact with the virtual CMDP to learn constraint-satisfying policies, avoiding safety risks in the early exploration. The pretrained policies are subsequently fine-tuned via limited interactions with the real environment to obtain the optimal policy under the current environmental conditions.


In the following, we first discuss the behavior policy selection. Then, we present the implementation of the three key elements in CMDP modeling, i.e., reward and cost functions, initial-state distribution, and state transition probability. These components correspond to three distinct learning tasks: deterministic value prediction, distribution modeling, and conditional distribution modeling. 

\subsection{Offline Dataset Construction}
The offline pretraining framework provides a safe and reliable environment for DRL exploration, which benefits from the virtual CMDP established on the offline dataset.
A randomized behavioral policy $\hat{\pi}^*\left(\boldsymbol{a}_t\mid\boldsymbol{s}_t\right)$ that uniformly samples actions from the action space $\mathbb{A}$ is employed. This design choice benefits the effective implementation of the virtual CMDP for the following reasons:
\begin{enumerate}
	\item{\textit{Value Range Coverage}: The interactions contain sufficient coverage of reward and cost values across different magnitudes, which helps alleviate systematic prediction errors (overestimation/underestimation) in the reward and cost prediction.}
	\item{\textit{Exploration Completeness}: The uniform sampling mechanism ensures adequate coverage of the state-action space, making it possible to solve the inevitable distribution shift problem in state transition modeling.}
\end{enumerate}
Notably, the initial-state distribution $\rho(\boldsymbol{s}_0)$ is decoupled from behavioral policy $\hat{\pi}\left(\boldsymbol{a}_t\mid\boldsymbol{s}_t\right)$, inherently preserving the consistency between offline and online initial-state distributions.

\subsection{Deterministic Prediction of Reward and Cost Functions}
The reward function $r(\boldsymbol{s},\boldsymbol{a})$ and cost function 
$c(\boldsymbol{s},\boldsymbol{a})$ are deterministic mappings from state-action pairs to scalar values. While a traditional approach employs multilayer perceptron (MLP) based on the universal approximation theorem to fit the function through supervised learning, Kolmogorov-Arnold network (KAN)\cite{liu2025kankolmogorovarnoldnetworks} inspired by Kolmogorov-Arnold representation theorem shows better parameter efficiency and robustness in certain cases\cite{liu2025kankolmogorovarnoldnetworks,10752997,10839363}, providing a viable alternative.

The Kolmogorov-Arnold representation theorem \cite{braun2009constructive} states that any multivariate continuous function $f(\boldsymbol{x})$ can be expressed as a finite composition of univariate functions:
\begin{equation}\label{KAtheorem}
	f(\boldsymbol{x})=\sum\nolimits_{q=1}^{2n+1}\Phi_q\left(\sum\nolimits_{p=1}^n\phi_{q,p}(x_p)\right),
\end{equation}
where $\Phi_q$ and $\phi_{q,p}$ are univariate functions. 

\begin{figure}
	\centering
	\includegraphics[height=2.2in]{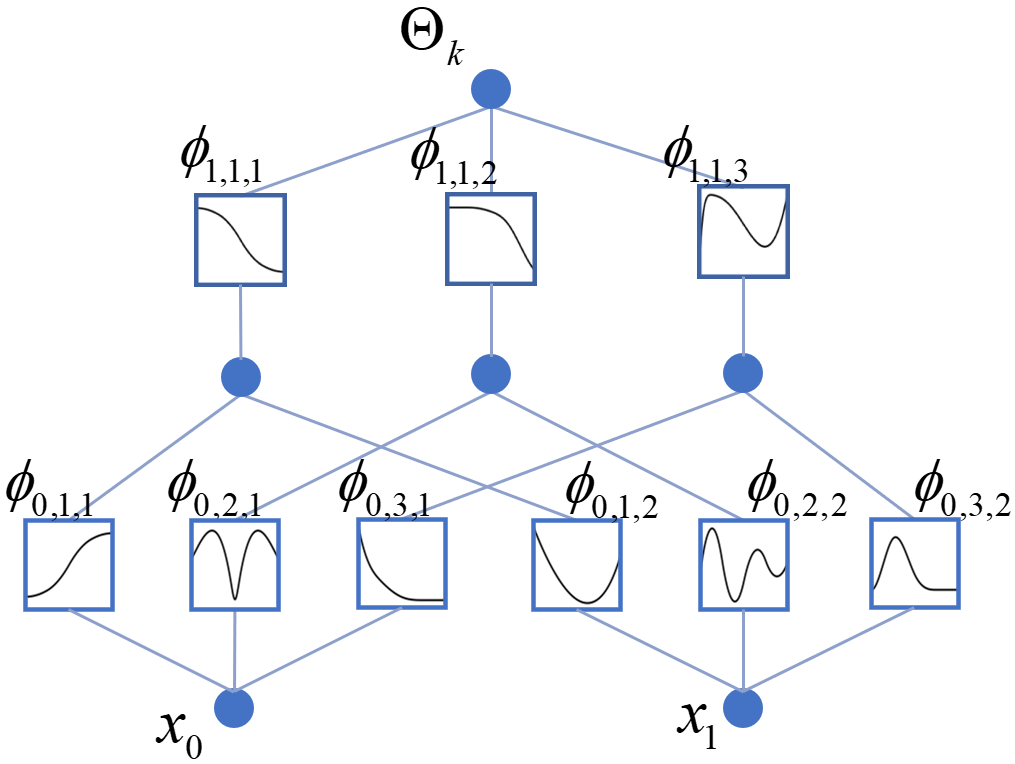}
	\caption{Illustration of a two-layer KAN architecture with learnable edge activation functions.}
	\label{KANfigure}
\end{figure}

KAN leverages neural networks to extend Eq. \eqref{KAtheorem} to arbitrary widths and depths. This extension addresses practical limitations in learning non-smooth $\Phi_q$ and $\phi_{q,p}$. For an $L$-Layer KAN, let $n_i$ represent the number of nodes in the $i$-th layer. Between the $l$-th and $(l+1)$-th layers, $n_{l}n_{l+1}$ activation functions are defined as $\phi_{l,j,i}$ for $l = 0, \ldots, L-1$, $j = 1, \ldots, n_{l+1}$ and $i = 1, \ldots, n_l$.
Given an input $\boldsymbol{x} \in \mathbb{R}^{n_0}$, the output of an $L$-layer KAN is formulated as 
\begin{equation}
	\text{KAN}(\boldsymbol{x}) = \sum\nolimits_{i_{L-1}=1}^{n_{L-1}} \phi_{L-1,i_L,i_{L-1}} \left( \Theta_{L-1} \right),
\end{equation}
where $\Theta_{k} = \sum_{i_{k-1}=1}^{n_{k-1}} \phi_{k-1,i_k,i_{k-1}} \left( \Theta_{k-1} \right) \text{ for } k \geq 2$ and 
$\Theta_0 = \sum_{i_0=1}^{n_0} \phi_{0,i_1,i_0}(x_{i_0})$.
The learnable univariate function $\phi(\cdot)$ comprises a basis function $b(x)$ and a spline function:
\begin{equation}
	\phi(x) = w_b b(x) + w_s \mathrm{spline}(x),
\end{equation}
where $b(x)$  is typically implemented as the sigmoid linear unit $x/(1+e^{-x})$. The spline function, defined on a grid with $g$ intervals, is parameterized as a linear combination of $d$-order B-splines:
\begin{equation}
	\mathrm{spline}(x) = \sum\nolimits_{i=1}^{d+g} c_i B_i(x),
\end{equation}
with $B_i(x)$ denoting the $i$-th B-spline basis function. The weights $w_b$, $w_s$ and $c_i$s are trainable parameters.
A simple two-layer KAN architecture is illustrated in Fig. \ref{KANfigure}.

Note that all operations in KAN are differentiable and the mean squared error (MSE) loss between the predicted reward/cost and corresponding ground truth can be applied to train the networks with back propagation.

\subsection{Generative Modeling of the Initial-State Distribution} \label{VAE-ChMDN}
The initial-state distribution $\rho$ defines the starting conditions of the CMDP. Accurately modeling $\rho$ is critical for pretraining agents in scenarios with multi-modal initial configurations.
Among conventional generative approaches, GAN suffers from training instability due to non-convergent Nash equilibria in minimax optimization and mode collapse caused by the generator's exploitation of discriminator blind spots. Moreover, VAE exhibits restricted mode diversity due to its KL-divergence constrained latent variable noise and MSE reconstruction loss. 

To address these limitations, we employ our Cholesky decomposition-based VAE-Mixture Density Network (VAE-ChMDN) \cite{VAEChMDN} that integrates VAE with MDN featuring full-covariance GMM modeling.
This hybrid architecture employs a novel Cholesky-decomposition based training mechanism that simultaneously ensures the validity of covariance matrix and numerical stability during training process, overcoming key challenges in multi-modal explicit distribution learning. 

\begin{figure}
	\centering
	\includegraphics[width=3.3in]{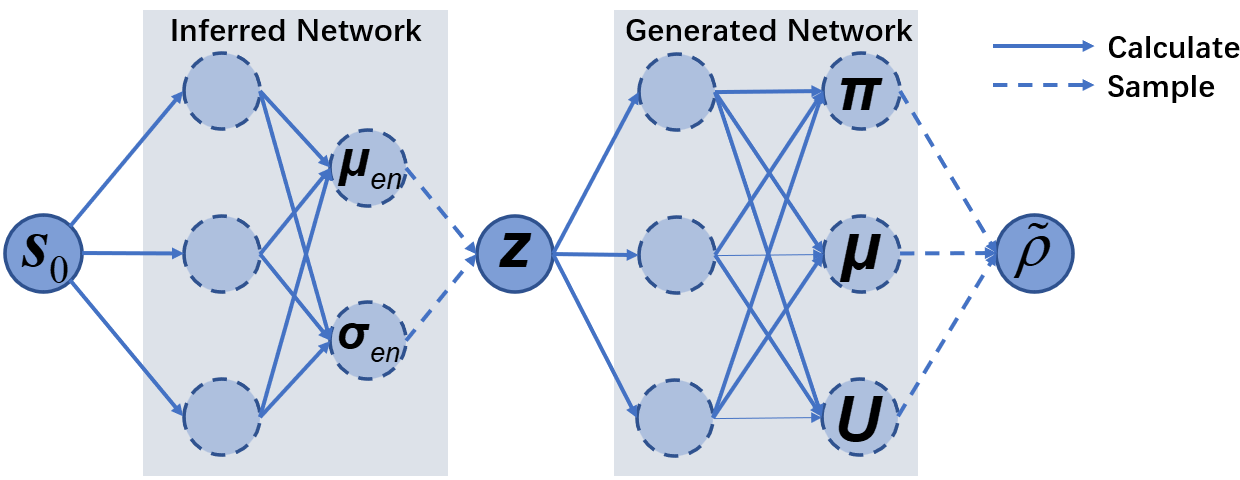}
	\caption{The structure of VAE-ChMDN.}
	\label{VAEMDNfigure}
\end{figure}

As shown in Fig. \ref{VAEMDNfigure}, the framework preserves the VAE's structured latent representation, where the inferred network constructs the latent variable $\boldsymbol{z} \sim \mathcal{N}(\boldsymbol{\mu}_{en},\boldsymbol{\sigma}_{en})$ as the potential expression of the input $\boldsymbol{s}_0$. The framework further leverages the MDN to output GMM parameters: mixture weights $\boldsymbol{\pi}=(\pi_1,\pi_2,\ldots,\pi_G)$, mean vectors $\boldsymbol{\mu}=(\boldsymbol{\mu}_1,\boldsymbol{\mu}_2,\ldots,\boldsymbol{\mu}_G)$, and Cholesky factors $\boldsymbol{U}=(\boldsymbol{U}_1,\boldsymbol{U}_2,\ldots,\boldsymbol{U}_G)$.

The initial-state distribution $\rho(\boldsymbol{s}_0)$ is modeled as a GMM with $G$ components:
\begin{equation}\label{GMMeq}
	\tilde{\rho}(\boldsymbol{s}_0)=\sum\nolimits_{g=1}^G\pi_gp_g(\boldsymbol{s}_0),
\end{equation}
where $\pi_g \in \left[0,1\right]$ satisfies $\sum_{g=1}^{G} \pi_{g}=1$, and $p_{g}(\boldsymbol{s}_0), g=1,\ldots,G$ is the probability density of the $g$-th multivariate Gaussian distribution:
\begin{equation}
	p_g(\boldsymbol{s}_0) \propto \exp\left(-\frac{1}{2}(\boldsymbol{s}_0 - \boldsymbol{\mu}_g)^\mathrm{T} \boldsymbol{\Sigma}_g^{-1} (\boldsymbol{s}_0 - \boldsymbol{\mu}_g)\right) \cdot |\boldsymbol{\Sigma}_g^{-1}|^\frac{1}{2}.
\end{equation}
Here, $\boldsymbol{\Sigma}_g=(\boldsymbol{U}_g^\mathrm{T}\boldsymbol{U}_g)^{-1}$ is the full-rank covariance matrix, where Cholesky factor $\boldsymbol{U}_g$ is an upper triangular matrix with positive diagonal elements, ensuring the symmetry and positive definiteness of $\boldsymbol{\Sigma}_g$.

The loss function consists of the KL divergence between the variational distribution $N(\boldsymbol{\mu}_{en},\boldsymbol{\sigma}_{en})$ and the prior $N(\boldsymbol{0},\boldsymbol{I})$ and the negative log-likelihood of the observed initial-state $\boldsymbol{s}_0$ under the generated GMM $\tilde{\rho}(\boldsymbol{s}_0)$, i.e.,
\begin{equation}
\text{Loss}=\mathrm{KL}(\mathcal{N}(\mathbf{0},\mathbf{I}),\mathcal{N}(\boldsymbol{\mu}_{\mathrm{en}},\boldsymbol{\sigma}_{\mathrm{en}}))-\operatorname{ln}\sum\nolimits_{g=1}^{G}\pi_{g}p_{g}(\boldsymbol{s}_0),
\end{equation}
where the log density of $\boldsymbol{s}_0$ for one mixture component in the GMM can be efficiently computed as $\ln p_g(\boldsymbol{s}_0)\propto-\frac{1}{2}\|\boldsymbol{U}_g(\boldsymbol{s}_0-\boldsymbol{\mu}_g)\|_2^2+\sum_{j=1}^{N}\ln \text{diag} \left(\boldsymbol{U}_g\right)_j$ following\cite{CholeskyMDNBase}.

When generating a GMM $\tilde{\rho}(\boldsymbol{s}_0)$, only the generated network is called, with its input $\boldsymbol{z}$ sampled from $\mathcal{N}(\boldsymbol{0}, \boldsymbol{I})$. 
During the sampling process for $\tilde{\rho}(\boldsymbol{s}_0)$, the $g$-th multivariate Gaussian distribution is sampled from the categorical distribution parameterized by $\boldsymbol{\pi}$, and a sample $\tilde{\boldsymbol{s}}_0^*$ is then obtained as follows:
\begin{equation}
	\tilde{\boldsymbol{s}}_0^*=\boldsymbol{\mu}_g+\boldsymbol{U}_g^{-1}\boldsymbol{\epsilon},
\end{equation}
where $\boldsymbol{\epsilon} \sim N(\boldsymbol{0},\boldsymbol{I})$ represents the sampling noise. The computational complexity of GMM sampling is $\mathcal{O}(\dim(\boldsymbol{s}_0)^3)$, primarily due to the matrix inversion of $\boldsymbol{U}_g$.

\subsection{Conditional Generative Modeling of the State Transition Probability}
Due to the uncertain channel state information (CSI) and packet arrivals at the next moment, the state transition probability $\mathcal{T}\left(\boldsymbol{s}_{t+1}\mid\boldsymbol{s}_t,\boldsymbol{a}_t\right)$ can be regarded as the distribution of $\boldsymbol{s}_{t+1}$ under conditions $\boldsymbol{s}_t$ and $\boldsymbol{a}_t$. 
While conventional conditional generative models, such as conditional VAE (CVAE) and conditional GAN (CGAN), perform well in discrete condition spaces through per-condition dense sampling, they face two fundamental limitations in continuous domains: 
\begin{enumerate}
	\item{\textit{Data sparsity}: For specific continuous condition values $(\boldsymbol{s}_t^*,\boldsymbol{a}_t^*)$, sufficient training samples are rarely available;}
	\item{\textit{Distribution-shift}: Models tend to overfit the offline behavioral policy's trajectory patterns, leading to extrapolation errors when simulating the agent policy's actions.}
\end{enumerate}

To address these challenges, we propose the EA-CGMM algorithm.
Specifically, the algorithm treats state transition tuple $(\boldsymbol{s}_{t+1},\boldsymbol{a}_t,\boldsymbol{s}_t)$ as a joint distribution, which is explicitly modeled as a $J$-component GMM $\{(\pi_{t,j}, p_j(\boldsymbol{s}_{t+1},\boldsymbol{a}_t,\boldsymbol{s}_t))\}_{j=1}^J$ through the VAE-ChMDN.
Leveraging the analytical properties of multivariate Gaussians derived in the Appendix, each multivariate Gaussian component $p_j(\boldsymbol{s}_{t+1},\boldsymbol{a}_t,\boldsymbol{s}_t)$ in the GMM can be decomposed into a marginal distribution $p_j(\boldsymbol{a}_t,\boldsymbol{s}_t)$ and a conditional distribution $p_j\left(\boldsymbol{s}_{t+1}\mid\boldsymbol{a}_t,\boldsymbol{s}_t\right)$. 
For the observed condition $(\boldsymbol{a}_t^*,\boldsymbol{s}_t^*)$, marginal distribution provides evidence weighting via $p_j(\boldsymbol{a}_t^*,\boldsymbol{s}_t^*)$, while conditional distribution represents state transitions  $p_j\left(\boldsymbol{s}_t \mid \boldsymbol{a}_t^*,\boldsymbol{s}_t^*\right)$, forming the conditional GMM framework.
The algorithm is detailed in the following.
\subsubsection{Explicitly Joint Distribution Modeling for Mitigating Data Sparsity}
In an MDP, state transitions satisfy the Markov property that the state of a future moment is only related to the state of its previous moment, i.e., $\boldsymbol{s}_{t+1}\perp\boldsymbol{s}_{1:t-1}|(\boldsymbol{s}_t,\boldsymbol{a}_t)$. Under time-homogeneous system dynamics, observed transitions form i.i.d. samples from stationary joint distribution $p(\boldsymbol{s}_{t+1},\boldsymbol{a}_t,\boldsymbol{s}_t)$.

To simplify notation, we slightly adjust the GMM parameter symbols from the VAE-ChMDN in Section \ref{VAE-ChMDN}, although the distribution modeled in this subsection differs from the previous one.
The proposed VAE-ChMDN models this joint distribution as a $J$-components GMM with parameters $\{(\pi_{j}, \boldsymbol{\mu}_{j}, \boldsymbol{U}_{j})\}_{j=1}^J$ through aggregated trajectory data, expressed as:
\begin{equation}
	\tilde{p}(\boldsymbol{s}_{t+1},\boldsymbol{a}_t,\boldsymbol{s}_t)=\sum\nolimits_{j=1}^J\pi_jp_j(\boldsymbol{s}_{t+1},\boldsymbol{a}_t,\boldsymbol{s}_t),
\end{equation}
where joint distribution of the state transition for the $j$-th component follows $(\boldsymbol{s}_{t+1},\boldsymbol{a}_t,\boldsymbol{s}_t)_j \sim \mathcal{N}(\boldsymbol{\mu}_{j}, (\boldsymbol{U}_{j}^\top\boldsymbol{U}_{j})^{-1})$.
Corresponding to $\boldsymbol{s}_{t+1}$ and $(\boldsymbol{a}_t,\boldsymbol{s}_t)$,  each joint mean vector $\boldsymbol{\mu}_{j}$ is partitioned into $\boldsymbol{\mu}_{j}^{(1)}$ and $\boldsymbol{\mu}_{j}^{(2)}$, and each joint Cholesky matrix $\boldsymbol{U}_{j}$ is partitioned into submatrices $\boldsymbol{U}_{j}^{11}$, $\boldsymbol{U}_{j}^{12}$, and $\boldsymbol{U}_{j}^{22}$ for the following derivation. To facilitate representation, let $\boldsymbol{w}_t$ denote the conditional tuple $(\boldsymbol{a}_t,\boldsymbol{s}_t)$.
According to Eq. $\eqref{aeq:cond}$ in the appendix, the conditional distribution of $j$-th component under observed value $\boldsymbol{w}_t^*=(\boldsymbol{s}_t^*,\boldsymbol{a}_t^*)$ is
\begin{equation}
	\left(\boldsymbol{s}_{t+1}\mid\boldsymbol{w}_t^*\right)_j\sim \mathcal{N}\left(\boldsymbol{\mu}_{j}^\prime, \boldsymbol{\Sigma}_{j}^\prime\right),
\end{equation}
with conditional mean and covariance satisfying
\begin{align}
	\boldsymbol{\mu}_{j}^\prime &= \boldsymbol{\mu}_{j}^{(1)} - \boldsymbol{U}_{j}^{11,-1}\boldsymbol{U}_{j}^{12}\left(\boldsymbol{w}_t^* - \boldsymbol{\mu}_{j}^{(2)}\right),\label{eq:cmean}\\
	\boldsymbol{\Sigma}_{j}^\prime &= \left( \boldsymbol{U}_{j}^{11,\top}\boldsymbol{U}_{j}^{11}\right)^{-1}.\label{eq:ccov}
\end{align}
This closed-form derivation enables robust density estimation by propagating global statistical patterns to local conditions, particularly effective under sparse observation regimes.

\subsubsection{Dynamic Reweighting Against Distribution Shift}
From Eq. \eqref{aeq:margin} in Appendix, the marginal distribution $p_j(\boldsymbol{w}_t)$ for component $j$ follows $\left(\boldsymbol{w}_t\right)_j \sim \mathcal{N}(\boldsymbol{\mu}_{j}^{(2)},(\boldsymbol{U}_{j}^{22,\top}\boldsymbol{U}_{j}^{22})^{-1})$. To obtain the updated mixture weight $\boldsymbol{\pi}_t^\prime$ of the conditional GMM, an intuitive idea is to use marginal probability density, i.e., $\pi_{j}^{\prime}=p_j({\boldsymbol{w}_t^*}),\forall j \in \{1,2,\ldots,J\}$. 
To improve robustness, we instead consider squared Mahalanobis distance, calculated as
\begin{equation}\label{Mahalanobis distance}
	\epsilon_j(\boldsymbol{w}_t^*)=\left(\boldsymbol{w}_t^*-\boldsymbol{\mu}_{j}^{(2)}\right)^{\top}\boldsymbol{U}_{j}^{22,\top}\boldsymbol{U}_{j}^{22}\left(\boldsymbol{w}_t^*-\boldsymbol{\mu}_{j}^{(2)}\right).
\end{equation}
The squared Mahalanobis distance eliminates the significant influence of covariance, reducing the situations that state transition is trapped in one component.
In order to adapt to the serious distribution-drift caused by the inconsistency between behavior policy $\hat{\pi}$ and agent policy, we propose a two-evidence-based reweighting scheme for mixture components. 

First, leveraging the fact that squared Mahalanobis distance follows a chi-squared distribution, we establish the upper $\left( 1-\alpha \right)$-quantile $\chi^2_{d, 1-\alpha}$ of the chi-squared distribution as the threshold for $\epsilon_j(\boldsymbol{w}_t^*)$, thereby constructing a statistical significance mask vector $\boldsymbol{m}=[m_1,m_2,\ldots,m_J]$. Each element of this binary vector is defined as:
\begin{equation}
	m_j = \begin{cases}
		1 & \text{if } \epsilon_j(\boldsymbol{w}_t^*) < \chi^2_{d, 1-\alpha} \\
		0 & \text{otherwise}
	\end{cases}, d=\dim(\boldsymbol{w}_t).
\end{equation}
Here, $m_j=1$ indicates that the corresponding conditional distribution $p_j\left(\boldsymbol{s}_{t+1}\mid\boldsymbol{w}_t^*\right)$ meets the credibility criterion. Conversely, $m_j=0$ signifies exclusion of the $j$-th mixture component from subsequent derivations.

Subsequently, the number of credible components is checked through $\boldsymbol{m}$. If $\|\boldsymbol{m}\|_1 > 0$, the updated weight $\boldsymbol{\pi}^\prime$ are uniformly distributed across all credible mixture components. Otherwise, $\boldsymbol{\pi}^\prime$ are assigned proportionally to the marginal probability densities $p_j(\boldsymbol{w}^*)$. This yields the update rule:
\begin{equation}
	\boldsymbol{\pi}^{\prime} = \begin{cases}
		\boldsymbol{m}/\|\boldsymbol{m}\|_1 & \text{if } \|\boldsymbol{m}\|_1 > 0 \\
		\mathrm{Softmax}\left(\left\{\ln p_j\left(\boldsymbol{w}^*\right)\right\}_{j=1}^J\right) & \text{otherwise}
	\end{cases}.
\end{equation}
The threshold $\chi^2_{d, 1-\alpha}$ for normal residuals establishes component-wise reliability, avoiding over-reliance on high-density components. When no components satisfy this criterion, which is common in high-dimensional spaces due to the curse of dimensionality, we revert to the marginal probability densities as weights, which is more sensitive to the covariance.

\begin{figure*}[t]
	\centering
	\includegraphics[width=\textwidth]{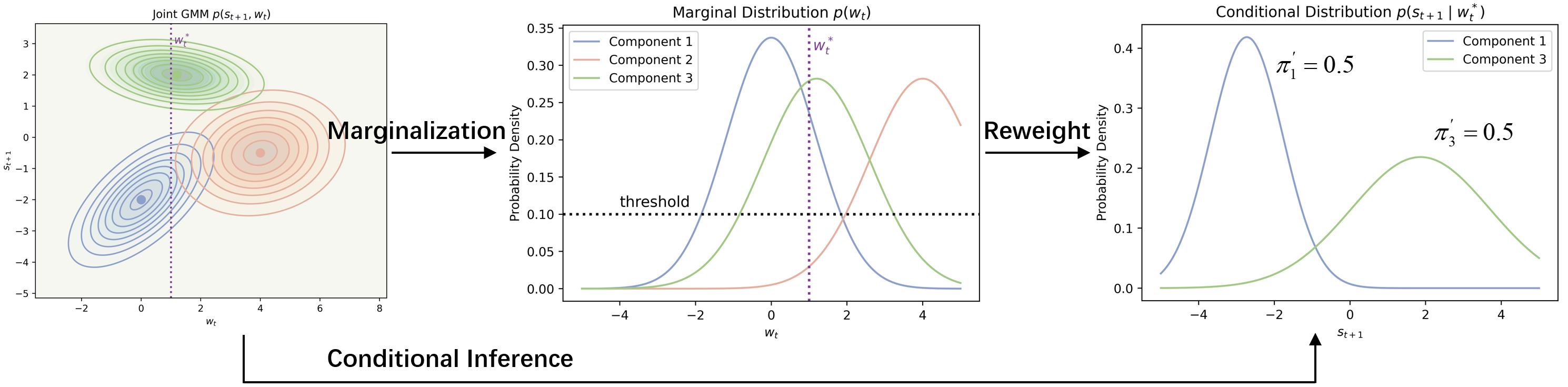}
	\caption{
		The joint distribution over $(\boldsymbol{s}_{t+1}, \boldsymbol{w}_t)$ is first modeled by a three-component GMM. After marginalizing $\boldsymbol{w}_t$, component-wise density evaluation is performed. Using the Mahalanobis distance in Eq.~(\ref{Mahalanobis distance}) together with the chi-squared distribution property, unreliable components (e.g., component $2$) are filtered out, and the mixture weights are uniformly redistributed among the remaining components. The conditional distribution $p(\boldsymbol{s}_{t+1} \mid \boldsymbol{w}_t^*)$ is then obtained in closed form using Eq.~(\ref{aeq:cond}).}
	\label{fig:ea_cgmm}
\end{figure*}

This produces a $J$-components conditional GMM $\left\{\left(\pi_{j}^{\prime}, p_j\left(\boldsymbol{s}_{t+1}\mid\boldsymbol{a}_t^*,\boldsymbol{s}_t^*\right)\right)\right\}_{j=1}^J$, with the computational complexity of the conditional GMM inference process being $\mathcal{O}(J\cdot\dim(\boldsymbol{s}_{t+1})^3)$. This complexity primarily attributed to matrix inversion operations involving $\boldsymbol{U}_{j}^{11}$ for $j=1,2,\ldots,J$.
The sampling process for $\left\{\left(\pi_{j}^{\prime}, p_j\left(\boldsymbol{s}_{t+1}\mid\boldsymbol{a}_t^*,\boldsymbol{s} _t^*\right)\right)\right\}_{j=1}^J$ and the corresponding computational complexity are detailed in Section \ref{VAE-ChMDN}.
Moreover, maintaining floating-point precision in high-dimensional matrix operations poses an additional computational challenge for both VAE-MDN and EA-CGMM.
The EA-CGMM algorithm is summarized in Algorithm \ref{alg:conditional_gmm}, while Fig.~\ref{fig:ea_cgmm} provides a visual illustration of overall procedure.
\begin{algorithm}
	\caption{Evidence-Aware Conditional GMM Inference}
	\label{alg:conditional_gmm}
	\renewcommand{\algorithmicrequire}{\textbf{Input:}}
	\renewcommand{\algorithmicensure}{\textbf{Output:}}
	\begin{algorithmic}[1]
		\REQUIRE Target condition $\boldsymbol{w}_t^*=\left(\boldsymbol{s}_t^*,\boldsymbol{a}_t^*\right)$ with dimension $d$
		\\ Upper $\left( 1-\alpha \right)$-quantile $\chi^2_{d, 1-\alpha}$ of the chi-squared distribution with $d$ degrees of freedom
		\\GMM parameters $\left\{\left(\pi_{j}, \boldsymbol{\mu}_{j}^{(1)}, \boldsymbol{\mu}_{j}^{(2)}, \boldsymbol{U}_{j}^{11}, \boldsymbol{U}_{j}^{12}, \boldsymbol{U}_{j}^{22}\right)\right\}_{1=1}^J$
		\ENSURE Conditional GMM $\left\{\left(\pi_{j}^{\prime}, p_j\left(\boldsymbol{s}_{t+1}\mid\boldsymbol{w}_t^*\right)\right)\right\}_{j=1}^J$
		\STATE Initialize statistical significance mask $\boldsymbol{m} \gets \boldsymbol{0}_J$, squared Mahalanobis distance vector $\boldsymbol{\epsilon}^* \gets \boldsymbol{0}_J$
		
		\FOR{$j = 1$ \TO $J$}
		\STATE Calculate $\boldsymbol{\mu}_{j}^\prime$ according to Eq. \eqref{eq:cmean}
		\STATE Calculate $\boldsymbol{\Sigma}_{j}^\prime$ according to Eq. \eqref{eq:ccov}
		\STATE $p_j\left(\boldsymbol{s}_{t+1}\mid\boldsymbol{w}_t^*\right) \gets \mathcal{N}\left(\boldsymbol{\mu}_{j}^\prime, \boldsymbol{\Sigma}_{j}^\prime\right)$
		
		\STATE $\boldsymbol{z}_j \gets \boldsymbol{U}_{j}^{22}\left(\boldsymbol{w}_t^* - \boldsymbol{\mu}_{j}^{(2)}\right)$ 
		\STATE $\epsilon_j^*\left(\boldsymbol{w}_t^*\right) \gets \boldsymbol{z}_j^{\top}\boldsymbol{z}_j$
		\STATE $m_j \gets \mathbb{I}\left[\epsilon_j^* < \chi^2_{d, 1-\alpha}\right]$ 
		\ENDFOR
		
		\IF{$\sum_{j=1}^J m_j > 0$}
		\STATE $\boldsymbol{\pi}_t^\prime \gets \boldsymbol{m}/\|\boldsymbol{m}\|_1$ 
		\ELSE
		\FOR{$j = 1$ \TO $J$}
		\STATE $\ln p_j\left(\boldsymbol{w}_t^*\right) \gets -\frac{1}{2}\|\boldsymbol{z}_j\|_2^2 + \sum_{i=1}^d \ln\mathrm{diag}\left(\boldsymbol{U}_{t,j}^{22}\right)_i$
		\ENDFOR
		\STATE $\boldsymbol{\pi}_t^\prime \gets \mathrm{Softmax}\left(\{\ln p_j\left(\boldsymbol{w}_t^*\right)\}_{j=1}^J\right)$
		\ENDIF
		
		\RETURN $\left\{\left(\pi_{j}^{\prime}, p_j\left(\boldsymbol{s}_{t+1}\mid\boldsymbol{w}_t^*\right)\right)\right\}_{j=1}^J$
	\end{algorithmic}
\end{algorithm}

With the virtual CMDP fully specified, we now briefly discuss the deployment of the pretraining framework in practical CF-MIMO systems.
In practical CF-MIMO control loops, data collection can be readily accomplished through randomized behavioral policy sampling. The virtual CMDP training and DRL pretraining phases can therefore be conducted periodically in the background, for instance on a minute- or hour-level basis. The online fine-tuning phase of the DRL can be executed real-time or near-real-time.

\section{Simulation Results and Discussions}
\label{Expiriment}
This section first introduces the simulated CF-MIMO OFDM downlink system along with its parameter configuration. Then proposed three key module implementations in the virtual CMDP are evaluated respectively, comparing with the current SOTA and baseline approaches. Finally, the effectiveness of our proposed pretraining framework is validated by the comparison between converged pretraining DRL agent and the non-pretrained baseline.

\subsection{Simulation Setup}
\begin{figure}
	\centering
	\includegraphics[width=3.3in]{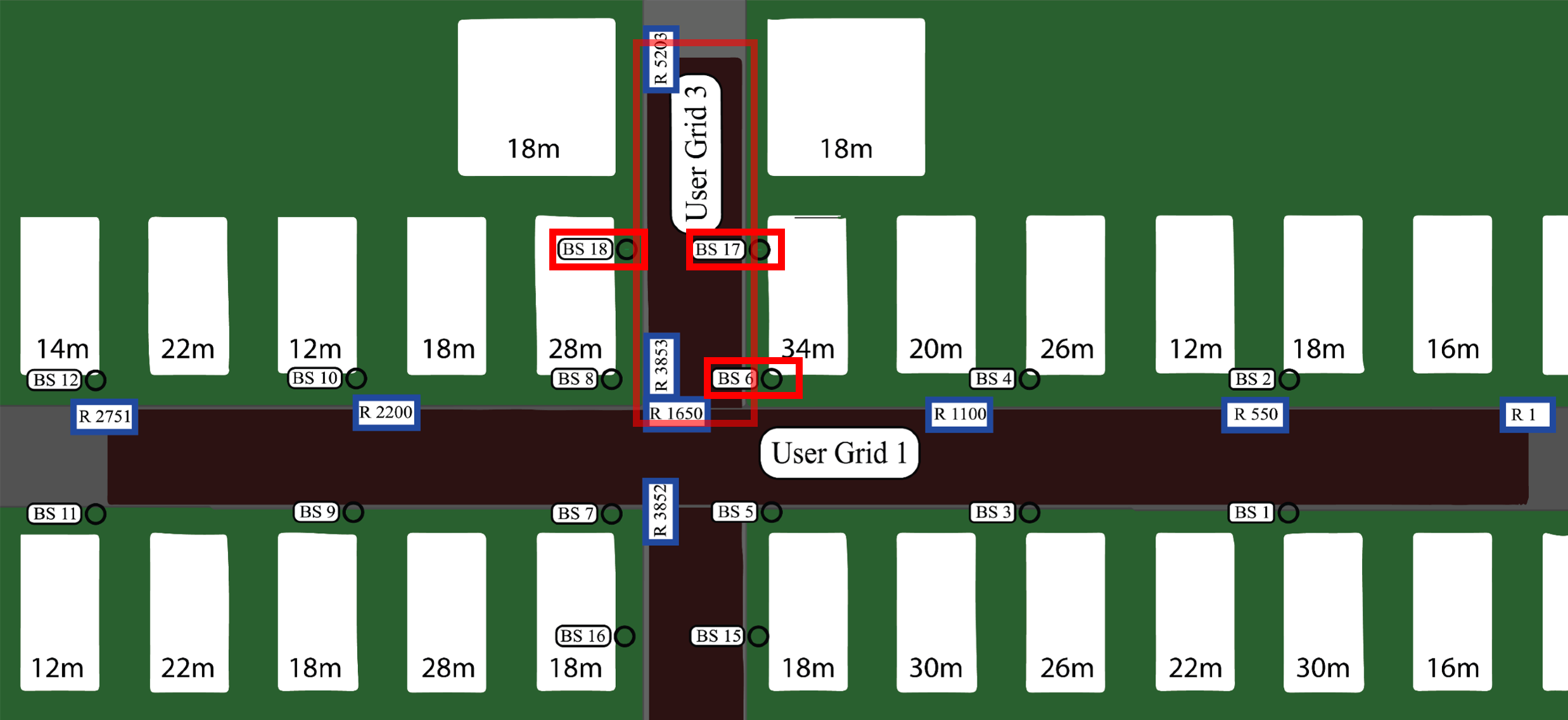}
	\caption{Partial top view of the “O1” scenario \cite{Alkhateeb2019}, showing selected APs $(6,17,18)$ and constrained UE mobility within User Grid $3$. The selected APs and User Grid constraint are marked within red boxes.}
	\label{O1figure}
\end{figure}

We consider a CF-MIMO OFDM downlink system, and the main simulation parameters are detailed in Table~\ref{Sim_para}.
Additionally, the time granularity is set as $\bigtriangleup t_s=5$ ms per slot and the system episode is set to $1$ s (equivalent to $200$ slots). OFDM system produces $CN=1500$ REs per minimum schedulable TFU, where $N=75$ corresponds to the number of OFDM symbols per slot derived from $\triangle t_s= N\frac{KC}{B_w}$.
\begin{table}
	\centering
	\caption{Key Simulation Parameters}
	\label{Sim_para}
	\footnotesize
	\begin{tabular}{|l|l|l|}
		\hline
		\textbf{Category} & \textbf{Parameter} & \textbf{Value} \\
		\hline
		\multirow{5}{*}{CF-MIMO}
		& Number of APs/UEs & $B=3$, $U=3$ \\ \cline{2-3}
		& Antennas per AP & $M=4$ \\ \cline{2-3}
		& Max transmit power & $P_{\max}=20$ dBm \\ \cline{2-3}
		& Noise PSD & $-174$ dBm/Hz \\ \cline{2-3}
		& Block error probability & $\epsilon_u=10^{-6}$ \\
		\hline
		\multirow{4}{*}{OFDM}
		& Bandwidth & $1.2$ MHz \\ \cline{2-3}
		& Subcarrier spacing & $15$ kHz \\ \cline{2-3}
		& Number of subbands & $K=4$ \\ \cline{2-3}
		& Subcarriers per subband & $C=20$ \\ 
		\hline
		\multirow{5}{*}{Per-user traffic}
		& Arrival process & Poisson, $\lambda_u=30$ \\ \cline{2-3}
		& Packet size & $\mathcal{U}[50,200]$ bits \\ \cline{2-3}
		& Latency constraint & $10$ ms \\ \cline{2-3}
		& Buffer size & $30$ kbits \\ \cline{2-3}
		& Scheduling & EDF-FIFO \\
		\hline
	\end{tabular}
\end{table}
The wireless propagation environment is generated using the O1 scenario from the DeepMIMO dataset \cite{Alkhateeb2019} with carrier frequency $3.4$ GHz, which provides ray-tracing-based CSI through Remcom Wireless InSite simulations \cite{Remcom}.
As shown in Fig. \ref{O1figure}, APs $6$, $17$, and $18$ are selected and each UE's position and CSI are bound to the predefined grid points in DeepMIMO O1 scenario. Specifically, the predefined grid points are composed of rows R$3860$ to R$5200$, with a spatial separation of $10$ cm between adjacent sampling points. Each UE is randomly assigned to an initial grid point with a predetermined movement direction. At every slot, users transition to adjacent grid points along their current direction. Upon reaching grid boundaries, they stochastically alter their movement direction while maintaining spatial adjacency constraints, continuing this pattern until the episode terminates. 



\begin{figure*}
	\centering
	\subfloat[]{\includegraphics[width=1.7in]{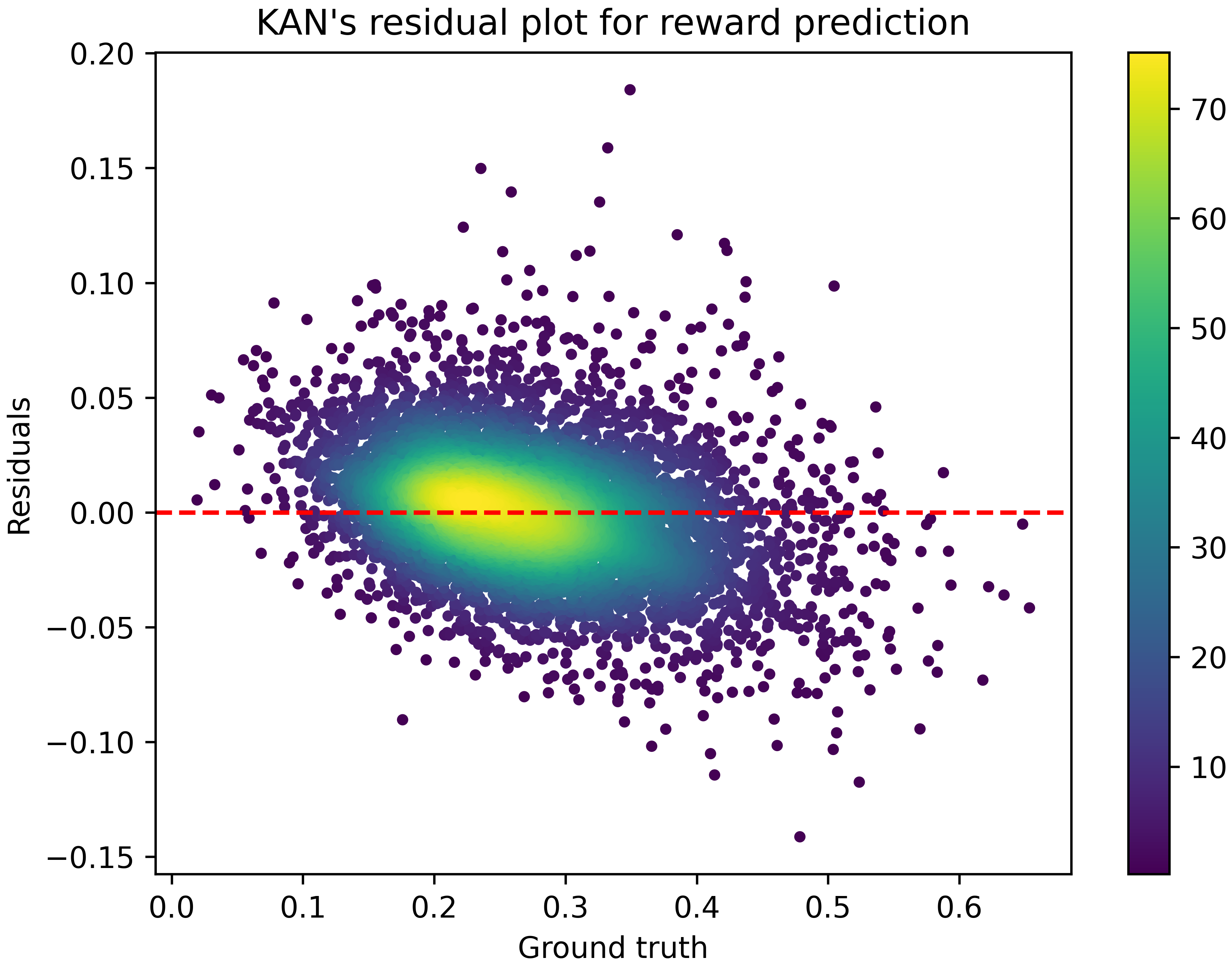}}%
	\label{subfigure_kanr}
	\hfil
	\subfloat[]{\includegraphics[width=1.7in]{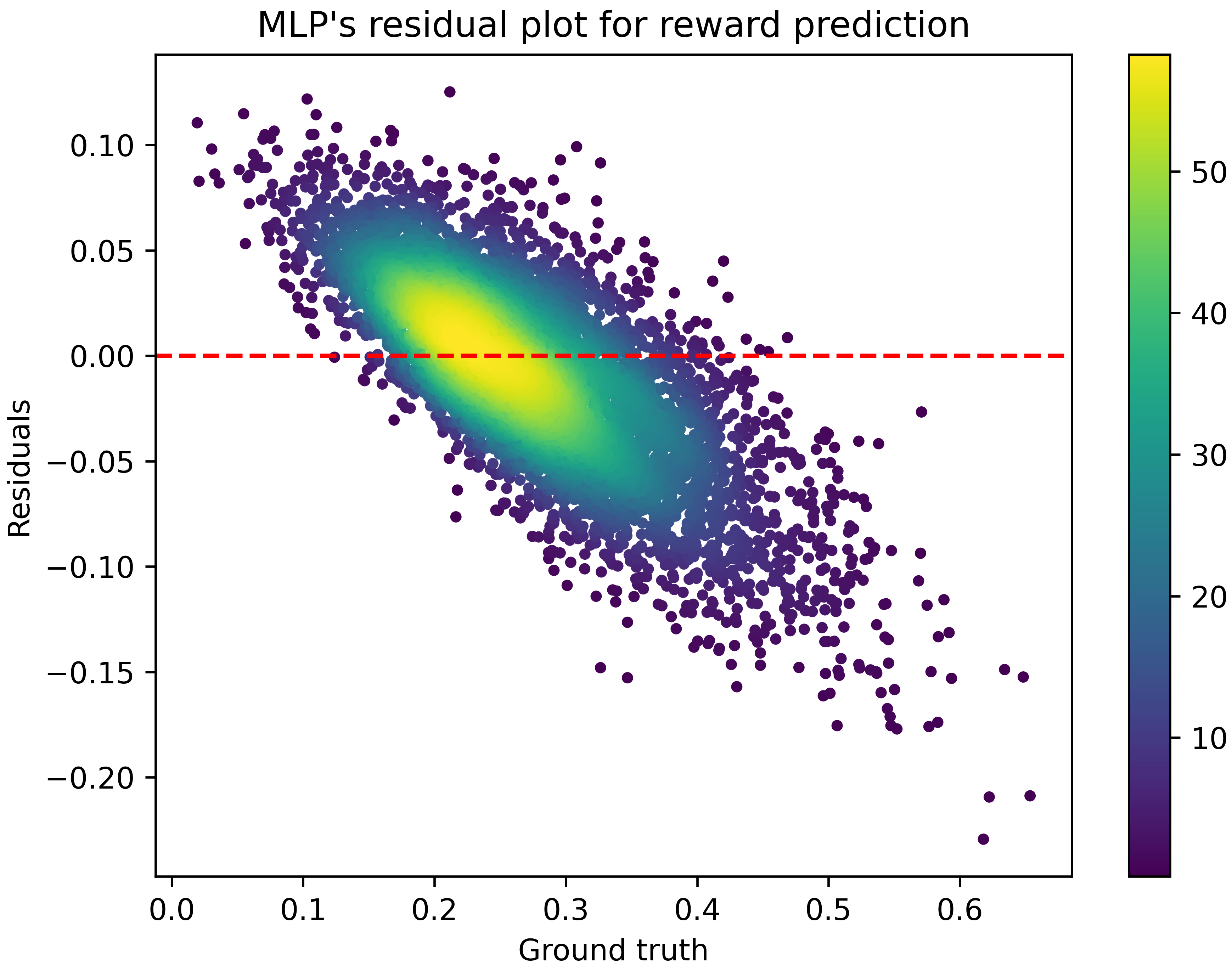}}%
	\label{subfigure_mlpr}
	\hfil
	\subfloat[]{\includegraphics[width=1.7in]{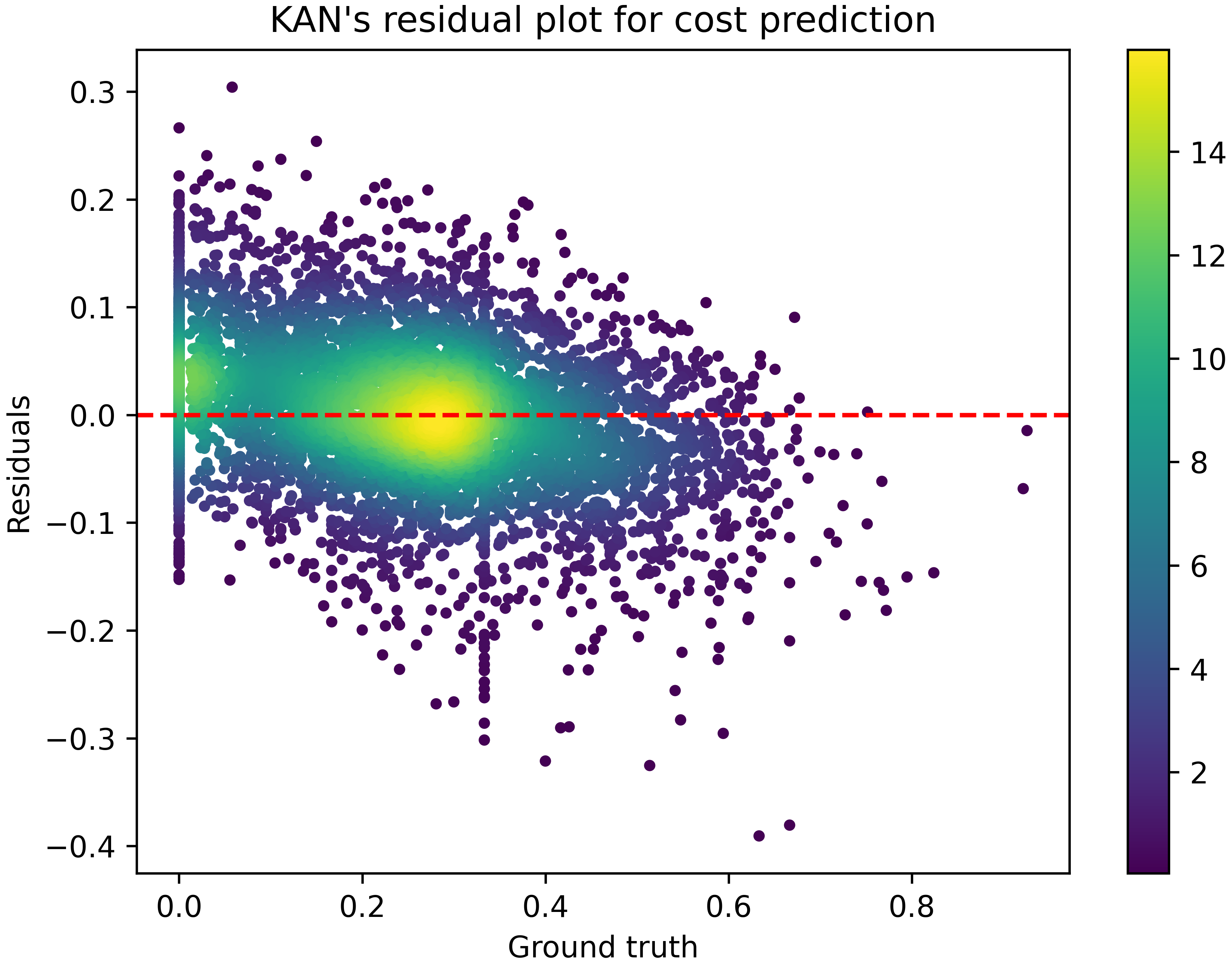}}%
	\label{subfigure_kanc}
	\hfil
	\subfloat[]{\includegraphics[width=1.7in]{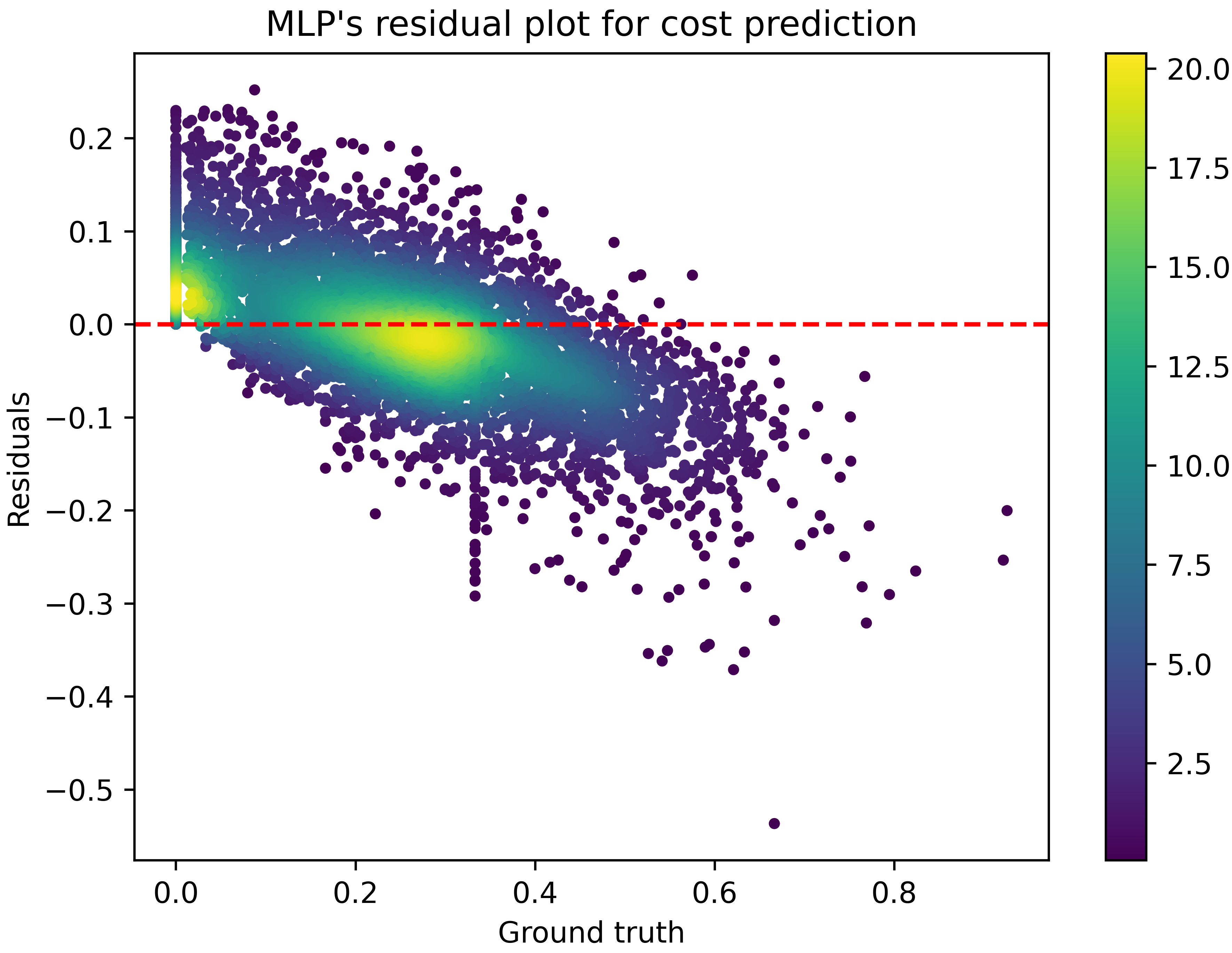}}%
	\label{subfigure_mlpc}
	\caption{Comparison of prediction residual distributions between KAN and MLP architectures, where horizontal axis represents ground truth and vertical axis denotes the residual between prediction value and the corresponding ground truth. (a) Reward residual of KAN, (b) Reward residual of MLP, (c) Cost residual of KAN, (d) Cost residual of MLP.}
	\label{residual_plot}
\end{figure*}

\subsection{Performance of Module Implementations in Virtual CMDP}\label{Performance_Modules}
In this subsection, our proposed virtual CMDP module implementations are evaluated and compared with corresponding SOTA and baseline methods.
Specifically, $30,000$ offline transition tuples $\left(\boldsymbol{s}_t,\boldsymbol{a}_t,r_t,c_t,\boldsymbol{s}_{t+1}\right)$ are collected through the randomized behavioral policy $\hat{\pi}^*$, which are randomly partitioned into an $80\%$ training set and $20\%$ test set.
In the reward and cost prediction module, KAN and MLP with comparable number of parameters are compared.
Regarding the initial-state modeling, VAE-ChMDN with mixture components $G=8$ is compared with recent SOTA denoising diffusion probabilistic model (DDPM)\cite{ho2020denoising} using $50$ diffusion steps and baseline generative models GAN and VAE. 
Building on these unconditional generative models, methods for state transition modeling introduce conditional mechanisms.
The proposed EA-CGMM approach with significance level $\alpha=0.03$ is compared with baselines CGAN and CVAE. Additionally, due to the frequent inference requirements, denoising diffusion implicit model (DDIM)\cite{song2020denoising} with classifier-free guidance\cite{ho2022classifier} using $5$ diffusion steps, that significantly accelerates the DDPM's diffusion process, is evaluated as the SOTA method.
The remaining experimental hyperparameters are listed in Table~\ref{hyper}.
\begin{table}[t]
	\centering
	\caption{Experimental hyperparameters}
	\label{hyper}
	\footnotesize
	\begin{tabular}{|l|p{3.5cm}|l|}
		\hline
		\textbf{Category} & \textbf{Description} & \textbf{Value} \\ 
		\hline
		
		\multirow{6}{2cm}{Training and evaluation setup} 
		& Optimizer & Adam \\ \cline{2-3}
		& Learning rate & $1\times10^{-4}$ \\ \cline{2-3}
		& Batch size & $256$ \\ \cline{2-3}
		& Random seed (data collection and models training) & $32$ \\ \cline{2-3}
		& Random seed (policy evolution evaluation) & $31$ \\ \cline{2-3}
		& Random seeds (pretraining performance evaluation) & $7-16$ \\ \hline
		
		\multirow{6}{2cm}{PPO hyperparameters}
		& Cost threshold & $d_u=0.5\%$ \\ \cline{2-3}
		& Initial dual variable & $\lambda=30$ \\ \cline{2-3}
		& Dual variable learning rate & $0.1$ \\ \cline{2-3}
		& Discount factor & $0.9$ \\ \cline{2-3}
		& Generalized advantage estimation parameter & $0.95$ \\ \cline{2-3}
		& Clip ratio & $0.2$ \\ \hline
		
		\multirow{3}{2cm}{KAN architecture} 
		& Grid size & 10 \\ \cline{2-3}
		& Spline order & 3 \\ \cline{2-3}
		& Grid epsilon & 0.1 \\ \hline
	\end{tabular}
\end{table}
\subsubsection{\bf Deterministic prediction performance}
Here, we compare the performance of KAN with $23,100$ parameters and MLP with $25,795$ parameters in the reward and cost prediction module. 
Experimental results demonstrate that KAN achieves superior performance with mean absolute errors (MAEs) of $0.0219$ (reward) and $0.0494$ (cost) on the test set, compared to MLP's MAEs of $0.0376$ and $0.0607$ respectively. 
Fig. \ref{residual_plot} further illustrates the residual distributions by plotting ground truth against prediction residuals. The residuals of KAN are nearly zero-mean with a slight inclination around the horizontal axis, indicating minor bias in estimation. In contrast, MLP displays systematic errors characterized by overestimation at lower target values and underestimation at higher magnitudes, suggesting suboptimal nonlinear fitting capability.
\subsubsection{\bf Initial-State Distribution Modeling Performance}
We employ the maximum mean discrepancy (MMD) for distribution similarity measurement. Two distributions are identical when MMD equals zero\cite{MMD2012}. Specifically, given $n$ samples $\{X_i\}_{i=1}^n \sim P(X)$, $m$ samples $\{Y_i\}_{i=1}^{m} \sim Q(Y)$, and the kernel function $k(\cdot)$, the empirical MMD is computed as:
\begin{equation}\begin{aligned}
		\widehat{\mathrm{MMD}}(P,Q) & =\frac{\sum_{i=1}^n\sum_{j=1}^nk\left(X_i,X_j\right)}{n^2} \\
		& +\frac{\sum_{i=1}^m\sum_{j=1}^mk\left(Y_i,Y_j\right)}{m^2} \\
		& -\frac{2\sum_{i=1}^n\sum_{j=1}^mk\left(X_i,Y_j\right)}{nm}.
\end{aligned}\end{equation}
Using the Gaussian kernel, we consider the case $m=n$ for MMD calculation.
Model complexity is quantified via floating point operations (FLOPs) per generated sample. 

Leveraging independence between PBM $\boldsymbol{r}_t$ and queue status information $\boldsymbol{q}_t$, $\rho(\boldsymbol{s}_0)$ is decomposed into $\rho(\boldsymbol{r}_0)$ and $\rho(\boldsymbol{q}_0)$ and modeled separately to reduce modeling complexity.
The models were trained using $120$ initial-state $\boldsymbol{s}_0$ samples extracted from the collected $30,000$ transition tuples $\left(\boldsymbol{s}_t,\boldsymbol{a}_t,r_t,c_t,\boldsymbol{s}_{t+1}\right)$ and evaluated on the remaining $30$ $\boldsymbol{s}_0$ samples.
As summarized in Table \ref{MMD_Comparison}, 
the VAE-ChMDN demonstrates comparable MMD to DDPM ($0.0022$ versus $0.0031$ for $\boldsymbol{r}_0$, and $0.0004$ versus $0.0014$ for $\boldsymbol{q}_0$), significantly outperforming GAN and VAE. 
\begin{table}
	\caption{MMD Comparison for Initial-State Distribution Modeling}
	\label{MMD_Comparison}
	\centering
	\begin{tabularx}{\linewidth}{lXXXX}
		\hline
		Model & $\text{MMD}_{\boldsymbol{r}}$ & $\text{FLOPs}_{\boldsymbol{r}}$& $\text{MMD}_{\boldsymbol{q}}$ & $\text{FLOPs}_{\boldsymbol{q}}$  \\
		\hline
		VAE-ChMDN    & $0.0022$ & $1.48$M &  $0.0004$   &  $0.12$M  \\
		DDPM  		 & $0.0031$ & $878.55$M &  $0.0014$   &  $98.40$M  \\
		GAN          & $0.0174$ & $0.77$M &  $0.0091$   &  $0.73$M  \\
		VAE          & $0.0293$ & $0.08$M &  $0.0057$   &  $0.06$M  \\
		\hline
	\end{tabularx}
\end{table}

Notably, the reported FLOPs for VAE-ChMDN correspond to the computational cost of both generating a GMM and drawing a sample from the resultant distribution, representing the upper-bound computational complexity per sampling operation.
While achieving similar modeling accuracy to DDPM, VAE-ChMDN reduces computational complexity by $590$-fold for $\boldsymbol{r}_0$ and $820$-fold for $\boldsymbol{q}_0$ in terms of FLOPs.
The computational complexity of VAE arises primarily from the Cholesky factors $\boldsymbol{U}$ that increases quadratically with the distribution dimension as well as the inversion operation for $\boldsymbol{U}$ when sampling from the GMM, and the complexity of DDPM comes from the repeated denoising over multiple diffusion steps through U-Net\cite{ronneberger2015u} architecture.
\begin{figure*}[ht]
	\centering
	\subfloat[]{
		\includegraphics[width=0.3\textwidth]{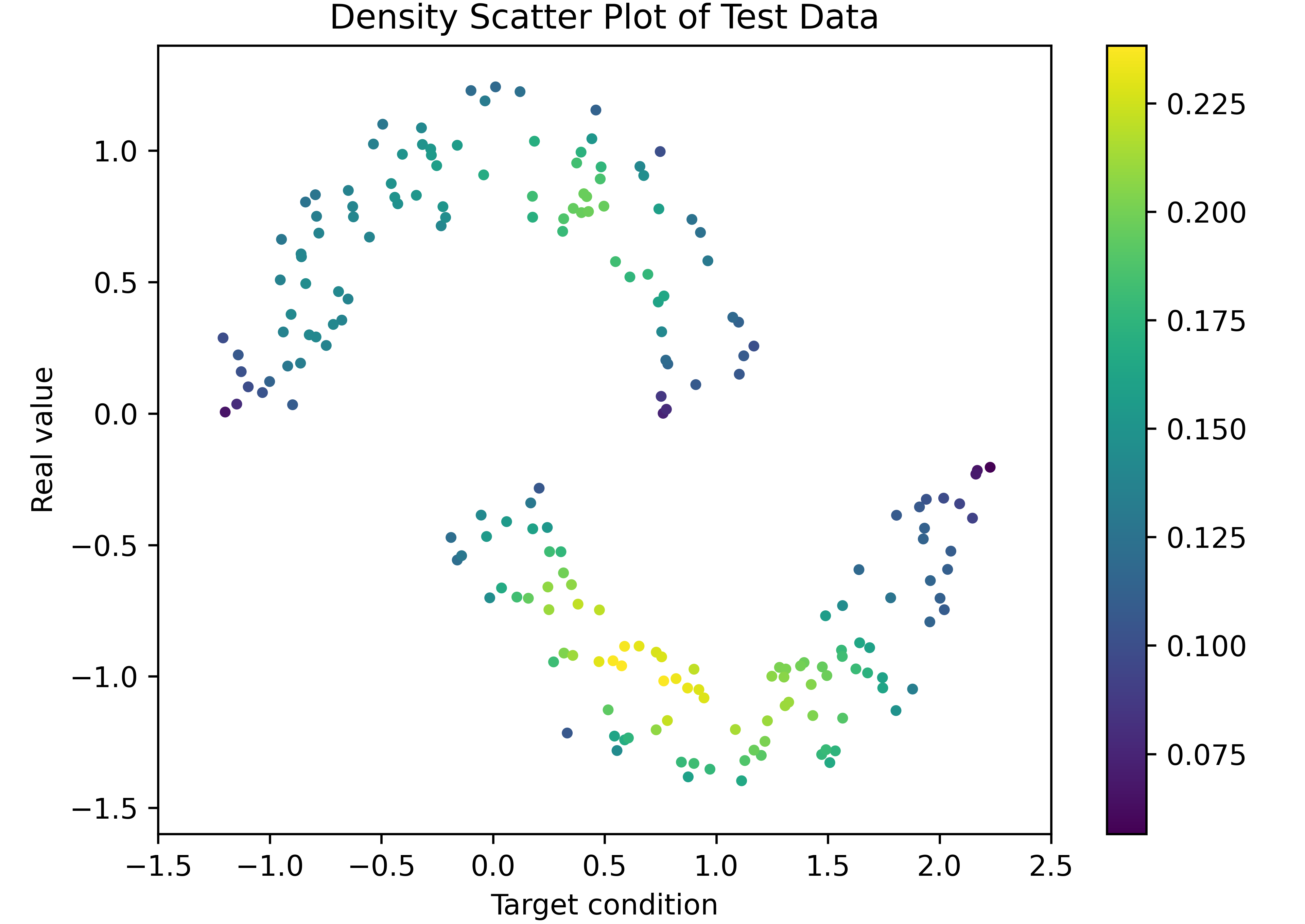}
		\label{toy_real}
	}
	\subfloat[]{
		\includegraphics[width=0.3\textwidth]{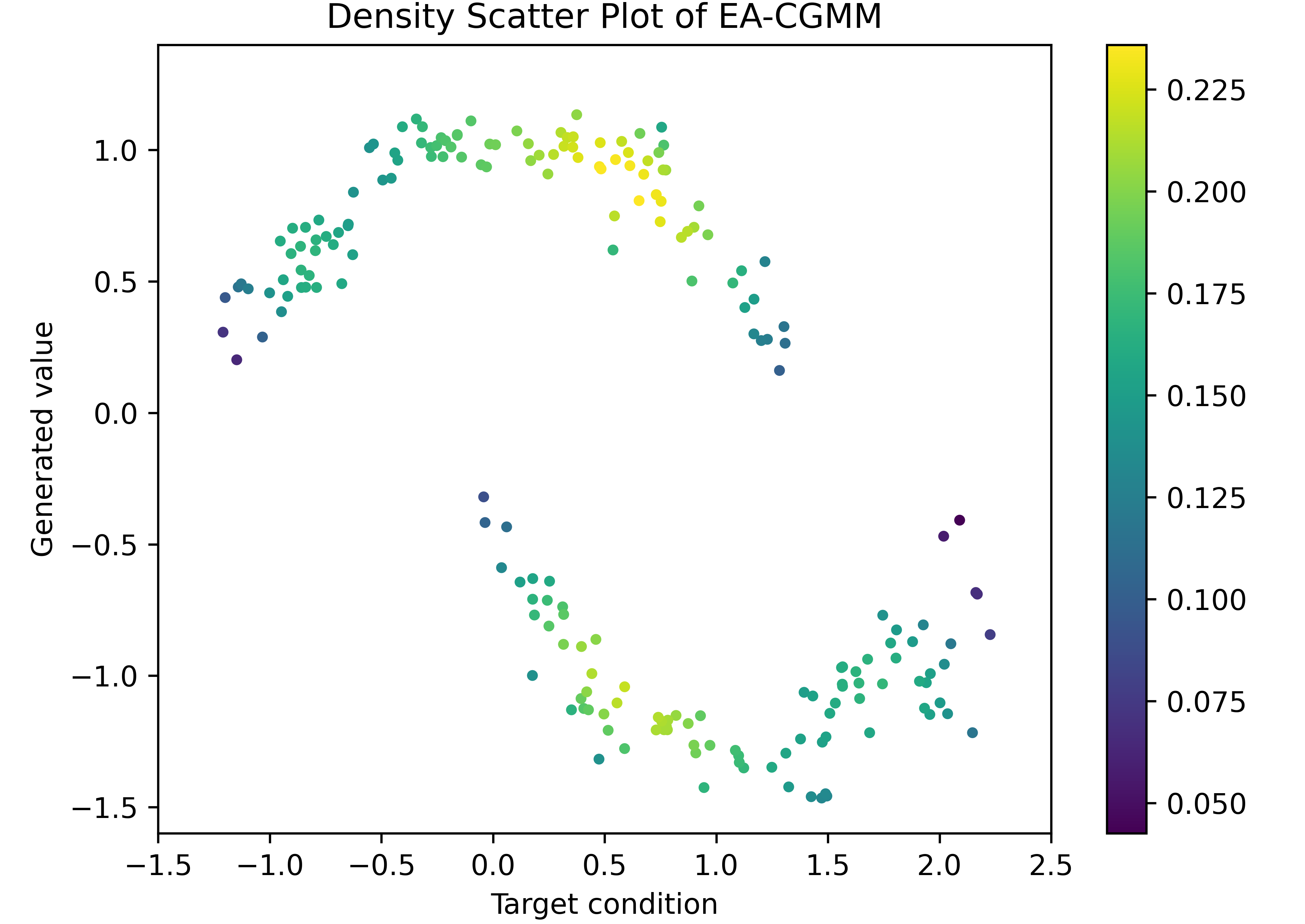}
		\label{toy_vaemdn}
	}
	\subfloat[]{
		\includegraphics[width=0.3\textwidth]{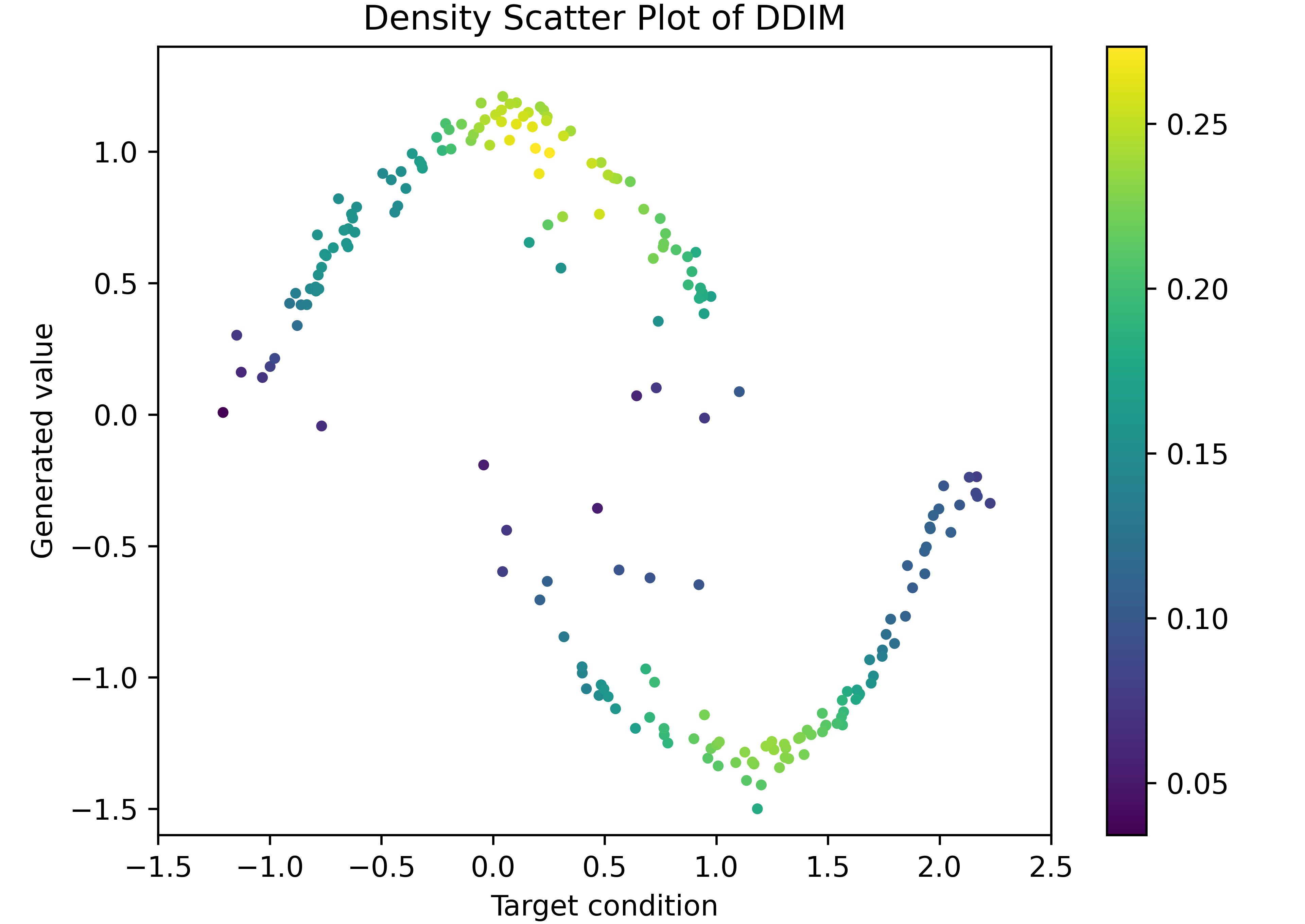}
		\label{toy_diff}
	}
	\vspace{-0.2cm}
	\\
	\subfloat[]{
		\includegraphics[width=0.3\textwidth]{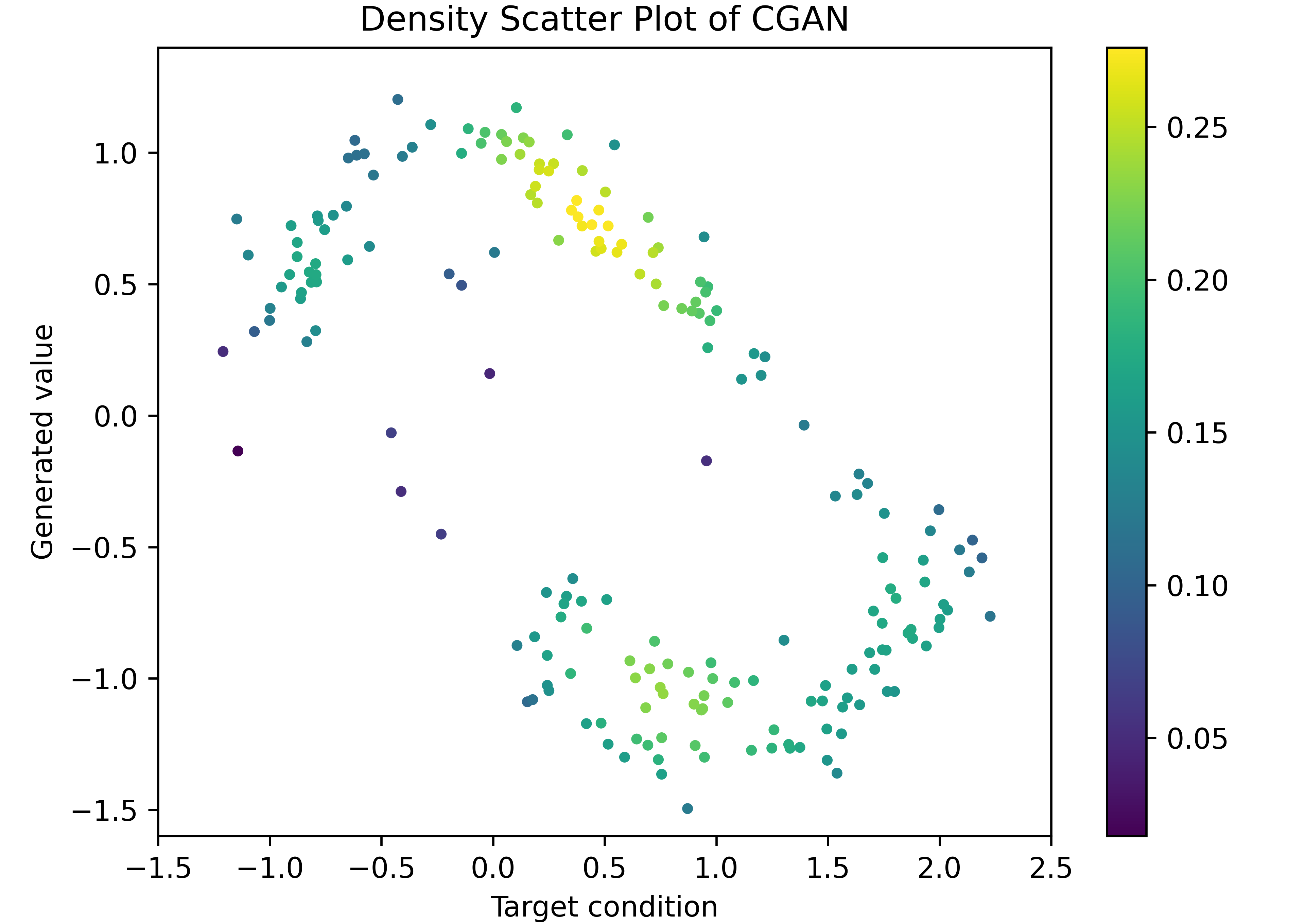}
		\label{toy_cgan}
	}
	\subfloat[]{
		\includegraphics[width=0.3\textwidth]{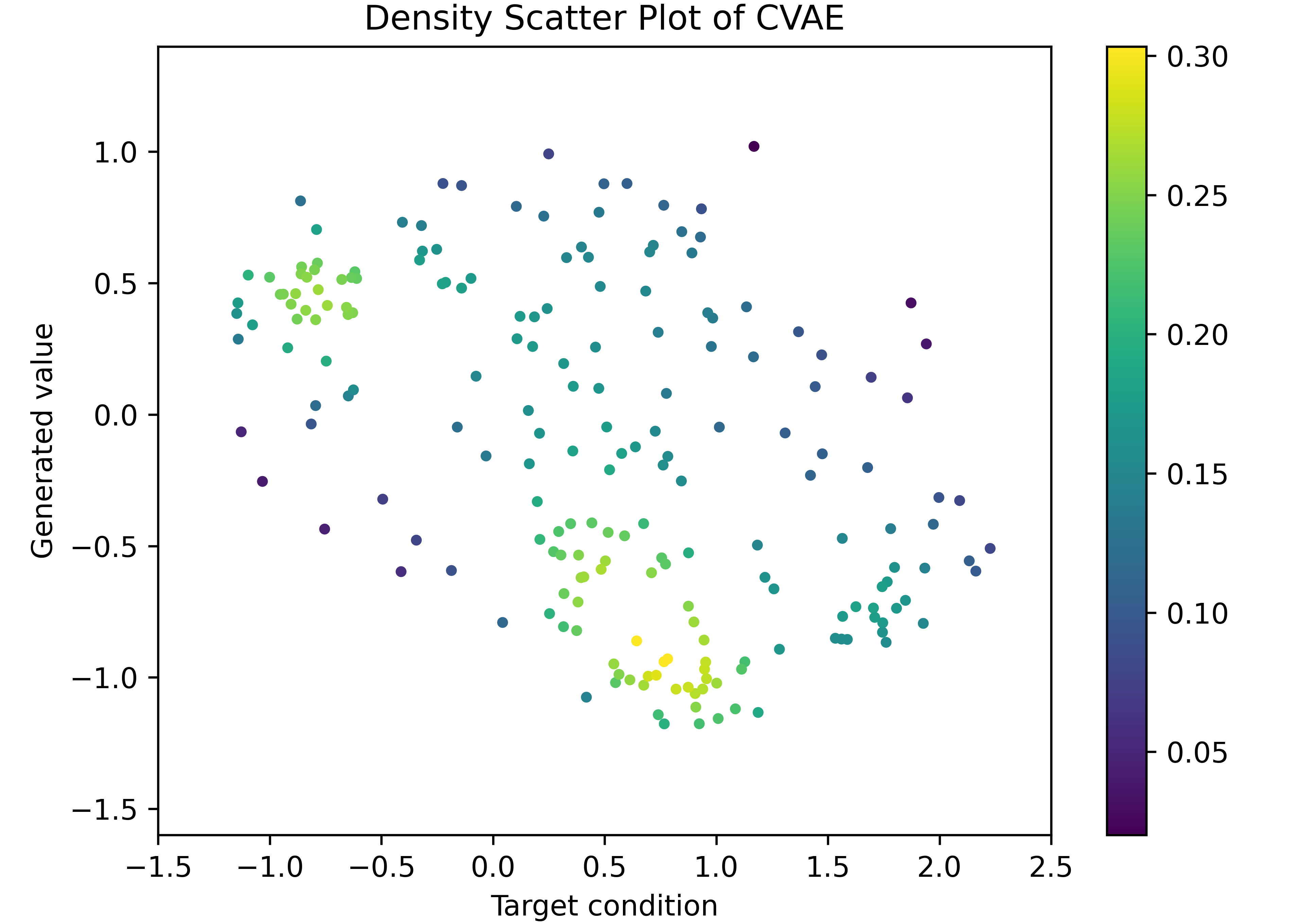}
		\label{toy_cvae}
	}
	\caption{Visualization of conditional generative modeling on a toy example using two half-moon benchmark, where horizontal axis represents target condition and vertical axis denotes target variable. (a) Test set distribution, (b) EA-CGMM samples, (c) DDIM samples, (d) CGAN samples, (e) CVAE samples.}
	\label{toy}
\end{figure*}
\subsubsection{\bf State Transition Modeling Performance}
Due to limited samples for MMD calculation under specific condition $(\boldsymbol{a}_t^*, \boldsymbol{s}_t^*)$, the MAE between generated $\left(\tilde{\boldsymbol{r}}_{t+1}^*,\tilde{\boldsymbol{q}}_{t+1}^*\right)$ and the corresponding ground truth $\left(\boldsymbol{r}_{t+1}^*,\boldsymbol{q}_{t+1}^*\right)$ is measured to evaluate model extrapolation ability in continuous condition space.

As shown in Table \ref{MAE_Comparison}, the EA-CGMM, DDIM, and CVAE demonstrate comparable MAE performance ($0.0289\textendash0.0459$ for $\boldsymbol{r}_{t+1}$ and $0.0486\textendash0.0793$ for $\boldsymbol{q}_{t+1}$), suggesting effective extrapolation capacities in continuous condition space. By contrast, CGAN exhibits significantly higher errors ($0.2101$ MAE for $\boldsymbol{r}_{t+1}$ and $0.2974$ MAE for $\boldsymbol{q}_{t+1}$), indicating limited extrapolation capability and suggesting mode collapse in high-dimensional spaces. 
\begin{table}
	\caption{MAE Comparison for Next-State Sample}
	\label{MAE_Comparison}
	\centering
	\begin{tabularx}{\linewidth}{lXXXX}
		\hline
		Method & $\text{MAE}_{\boldsymbol{r}}$ & $\text{FLOPs}_{\boldsymbol{r}}$ & $\text{MAE}_{\boldsymbol{q}}$ & $\text{FLOPs}_{\boldsymbol{q}}$ \\
		\hline
		EA-CGMM & $0.0459$ & $5.98$M & $0.0793$ & $0.83$M \\
		DDIM    & $0.0357$ & $87.86$M & $0.0486$ & $16.94$M \\
		CGAN    & $0.2101$ & $0.82$M & $0.2974$ & $0.75$M \\
		CVAE    & $0.0289$ & $0.08$M & $0.0652$ & $0.06$M \\
		\hline
	\end{tabularx}
\end{table}
In terms of computational complexity, the FLOPs quantification for EA-CGMM encompasses the total computational cost across three critical stages: GMM generation through VAE-ChMDN, conditional GMM inference for next-state, and probabilistic sampling from the derived conditional GMM. This aggregated measurement thereby defines the theoretical upper-bound complexity per sampling instance.
Although DDIM accelerates the diffusion process, it requires $14$ times and $20$ times more FLOPs than that of EA-CGMM for $\boldsymbol{r}_{t+1}$ and $\boldsymbol{q}_{t+1}$ generation, respectively.
Note that while the MAE provides practical insights for model's extrapolation ability, it does not fully characterize distributional fidelity due to the lack of generative diversity evaluation. 

In order to further reveal the efficiency of proposed EA-CGMM, as well as the limitations of the traditional models, we design a toy example using the well-known two half-moon rings benchmark for the conditional generative modeling task. Specifically, a $400$-sample two half-moon rings dataset is constructed with one dimension as the condition and the other as the target value. The dataset is divided into a train set of $200$ samples, and a test set of $200$ samples. As shown in Fig. \ref{toy}, both EA-CGMM and DDIM exhibit accurate inference while maintaining the diversity of generated samples, and the former represents more diversity while the latter achieves more accuracy in the trade-off. In contrast, CGAN maintains diversity, but the inference accuracy is lower. The generated distribution of CVAE is approximate to a joint Gaussian distribution, lacking of pattern diversity.

It should be noted that the FLOPs reported in Tables \ref{MMD_Comparison} and \ref{MAE_Comparison} correspond to the computational complexity of the virtual CMDP generation modules. These modules are used for constructing the virtual environment and are executed during the offline pretraining stage.
In the online deployment phase, the dominant computation stems from the action inference of the trained PPO policy. The inference requires approximately $20.6$k FLOPs, corresponding to an average latency of $76.9~\mu$s on an NVIDIA A100 GPU platform. Such latency is significantly lower than the millisecond-level timing requirement of URLLC systems, demonstrating that the proposed framework satisfies practical real-time constraints.

\subsection{Performance of Proposed Pretraining Framework}\label{Per_pretraining}
This subsection presents the experimental validation of our virtual CMDP-based pretraining framework. The evaluation maintains KAN for reward and cost prediction while implementing four comparative pretraining frameworks using the generative models and corresponding conditional generative modeling approaches outlined in Section \ref{Performance_Modules}. All frameworks employ PPO with a primal-dual approach for policy optimization.
The configuration of PPO is detailed in Table~\ref{hyper}, where the cost threshold $d_u=\eta_u=0.5\%, \forall u \in \mathbb{U}$ represents the target single-slot delay violation rate. The corresponding target delay violation probability $\eta_u$ in Constraint~\eqref{eq:constraint1} is empirically chosen as a feasible operating point to ensure stable training and satisfactory performance.
To better evaluate the pretraining performance, we also adopt Lyapunov optimization as a classical model-based baseline. Following the Lyapunov drift-plus-penalty framework\cite{neely2010stochastic}, at each time step we sample $200$ candidate actions and select the one minimizing a per-slot objective based on KAN’s predicted reward $\tilde{r}_t$ and cost $\tilde{c}_t$, where $-\tilde{r}_t$ is treated as the objective term and the deviation of $\tilde{c}_t$ from the threshold $d$ is incorporated via a virtual queue to enforce long-term constraint satisfaction.

\begin{figure}
	\centering
	\subfloat[]{\includegraphics[width=3in]{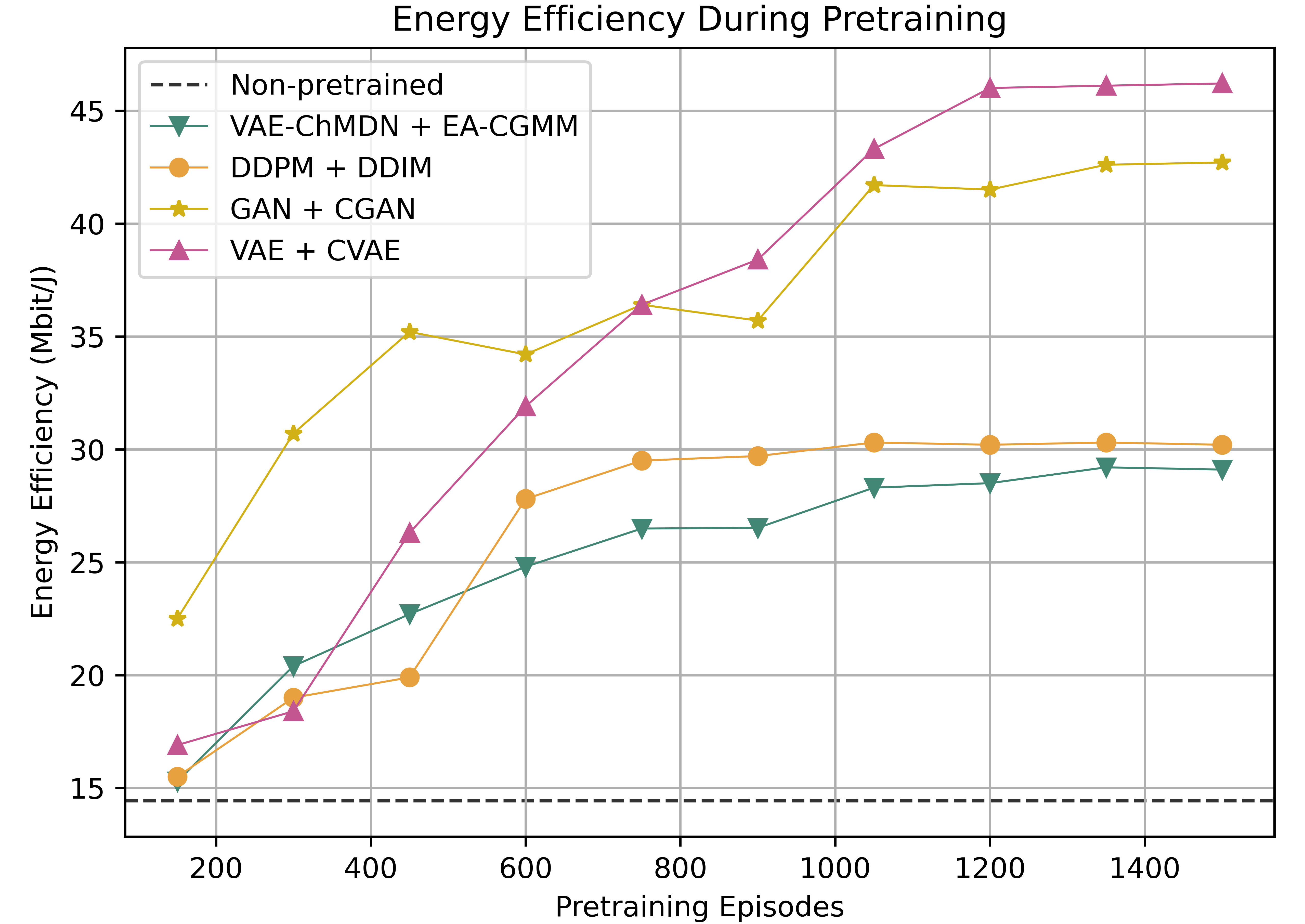}%
		\label{performance_pre_r}}\\
	\subfloat[]{\includegraphics[width=3in]{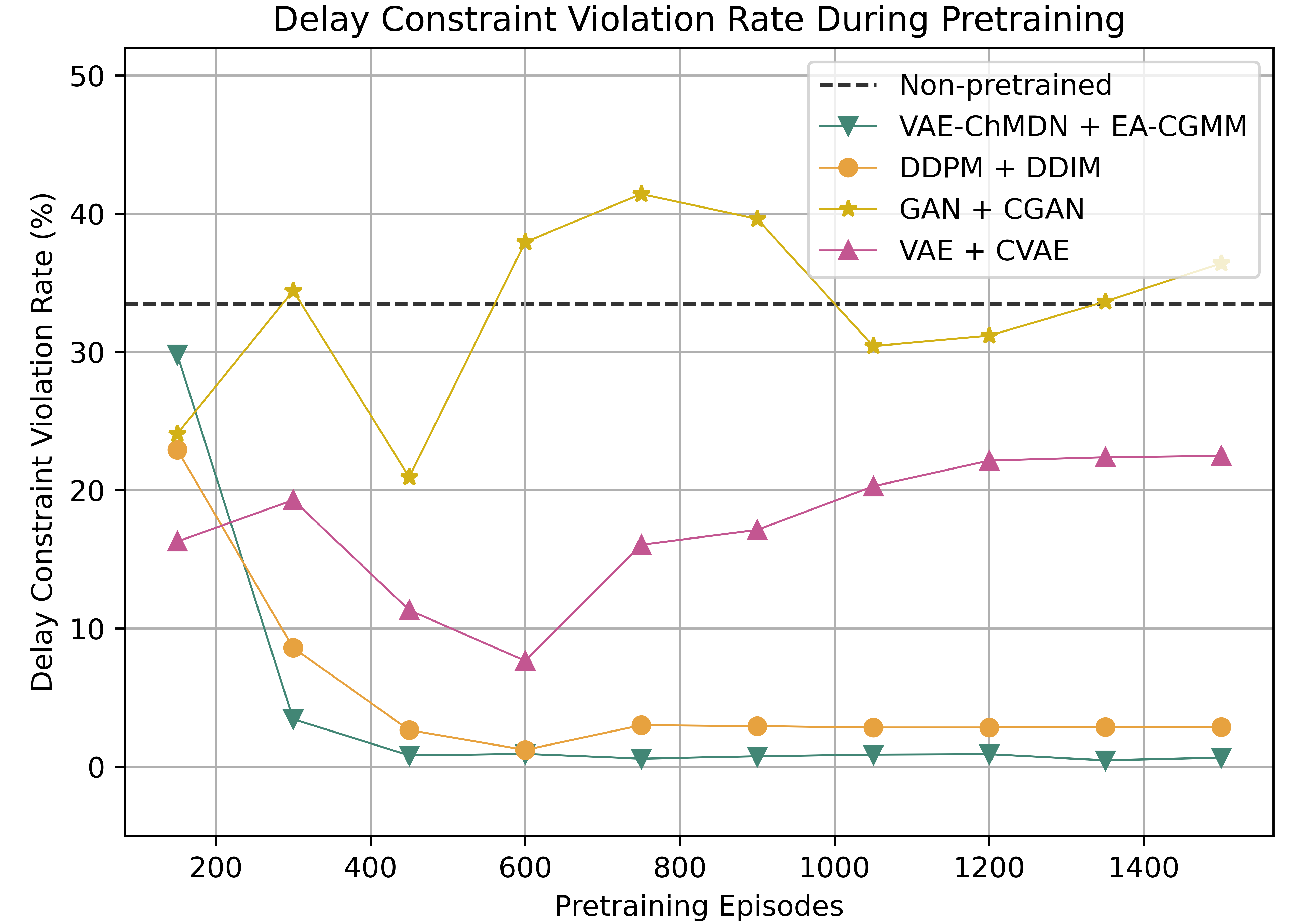}%
		\label{performance_pre_c}}
	\caption{Policy evolution during pretraining phase: (a) Average EE across different frameworks, (b) Corresponding delay constraint violation rates.}
	\label{Performance_pre}
\end{figure}

Fig. \ref{Performance_pre} illustrates the policy evolution during $1,500$ pretraining episodes, with policy snapshots evaluated through $100$ Monte Carlo trials per $150$-episode interval. 
The non-pretrained PPO baseline yields an initial EE of $14.4$ Mbit/J with a delay violation rate of $33\%$, serving as a reference for evaluating the effectiveness of the proposed pretraining frameworks.
Agents in the VAE-ChMDN-based and the DDPM-based pretraining frameworks achieve stable EE convergence at $29$ Mbit/J and $30$ Mbit/J respectively, with low initial delay violation rates of around $1\%$ and $3\%$, respectively.
Due to the substantial discrepancies between the VAE- and GAN-based frameworks and the real environment, neither curve demonstrates stable convergence during testing.
While the former framework achieves peak EE of $46$ Mbit/J, it exhibits a high delay constraint violation rate of near-$22\%$, indicating limited robustness from generative diversity constraints. The latter implementation shows degraded performance with $42$ Mbit/J EE and $35\%$ delay constraint violation rate, revealing poor extrapolation capability in policy guidance.

\begin{figure}
	\centering
	\subfloat[]{\includegraphics[width=3in]{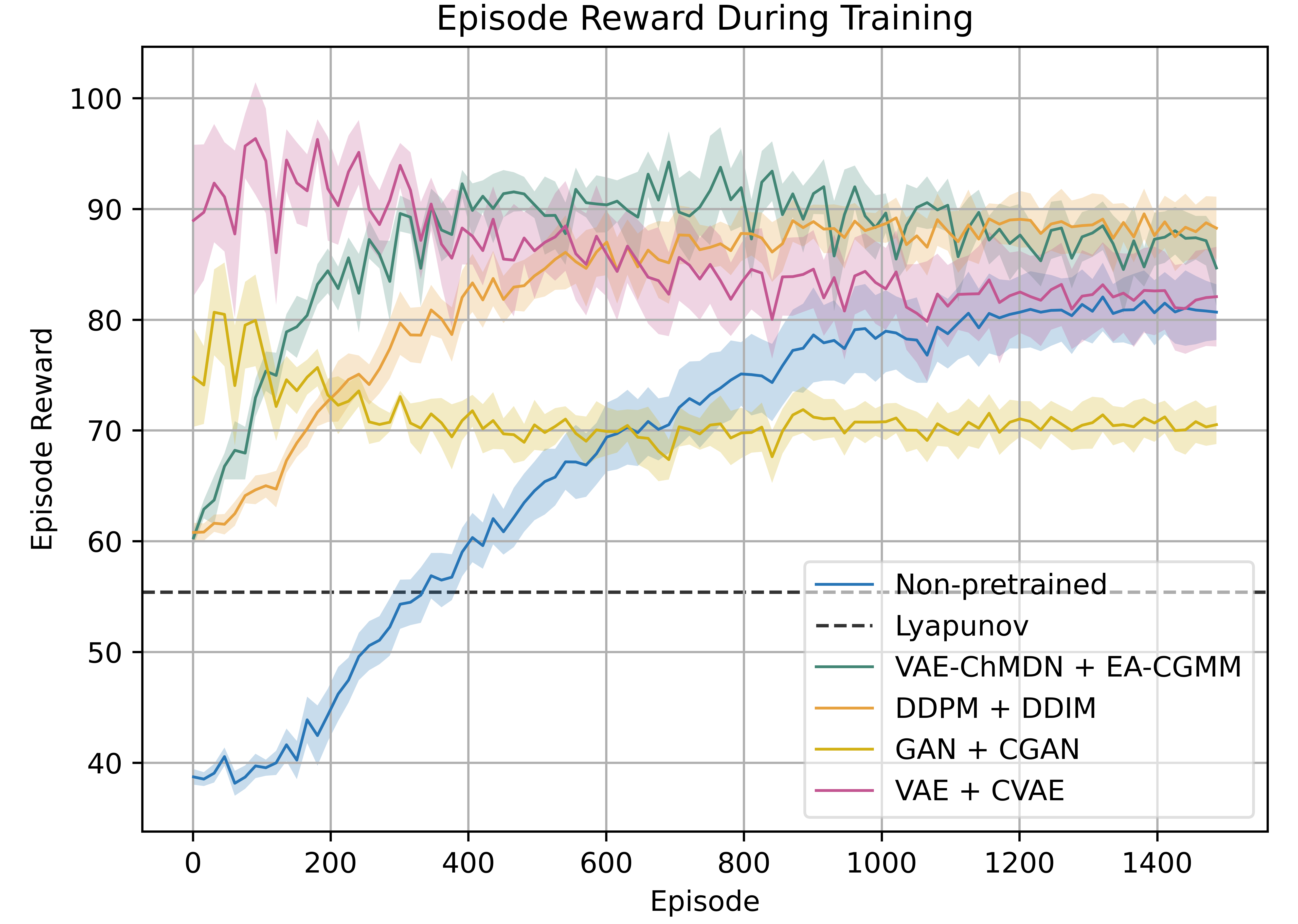}%
		\label{performance_pre_real_r}}\\
	\subfloat[]{\includegraphics[width=3in]{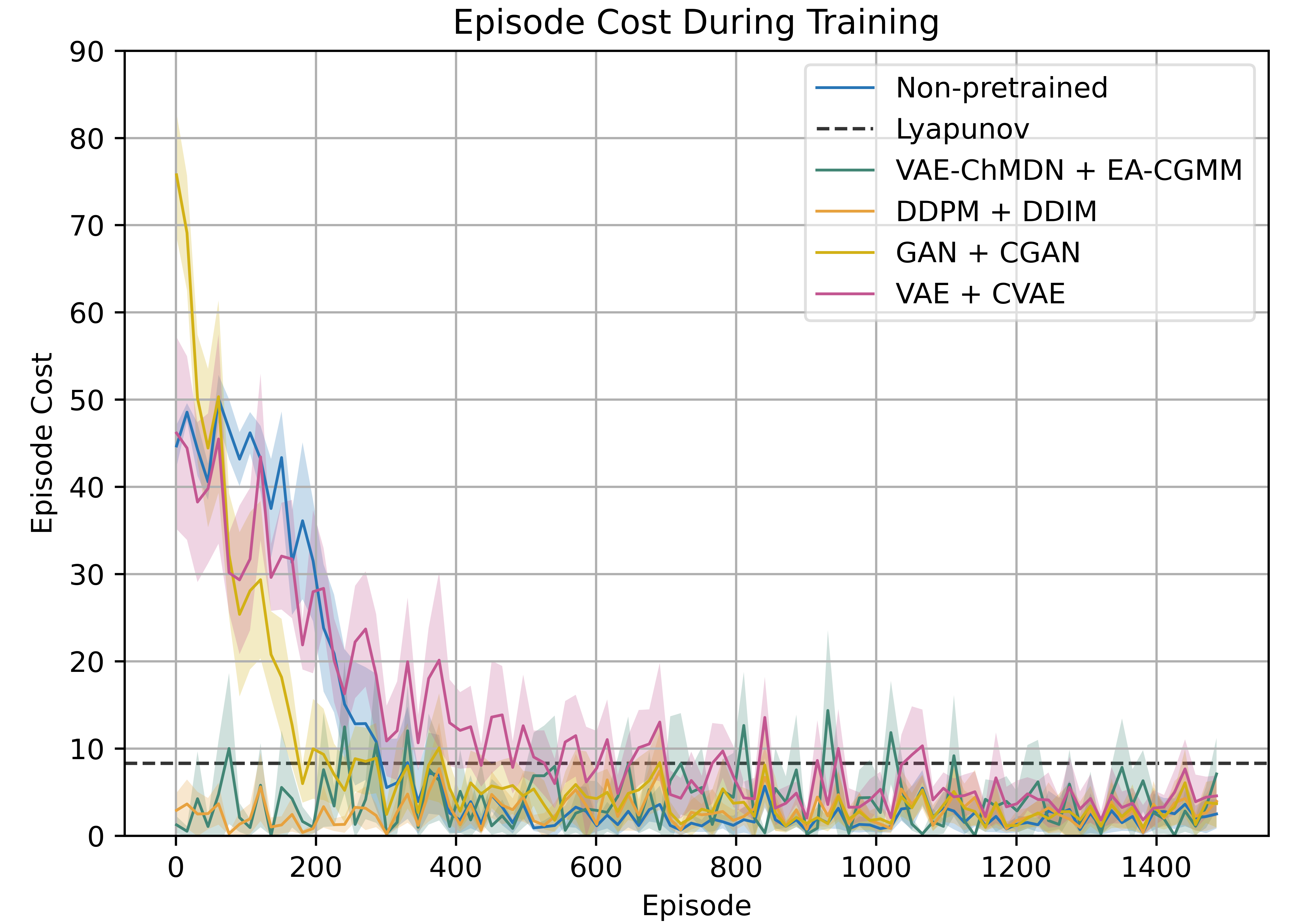}%
		\label{performance_pre_real_c}}
	\caption{Learning curves comparison of pretrained agents and non-pretrained baseline in simulation environment: (a) Episode reward evolution, (b) Corresponding episode cost evolution. Curves represent mean values with shaded regions denoting $\pm 0.5$ standard deviation across $8$ random seeds.}
	\label{Performance_pre_real}
\end{figure}

The fine-tuning efficiency of $1,200$-episode pretrained policies was further evaluated against non-pretrained baseline through $1,500$ simulation episodes with 8 random seeds. The Lyapunov baseline was evaluated by directly averaging its episode performance over the same $8$ random seeds, as it does not involve policy learning and applies a greedy deterministic optimization at each time step.
As depicted in Fig. \ref{Performance_pre_real}, the agent pretrained by VAE-ChMDN-based framework commences fine-tuning with a $52\%$ higher initial episode reward compared to the non-pretrained baseline and an initial episode cost that approximately satisfies the threshold. This agent converges $50\%$ faster ($600$-episode reduction) with $4.7\%$ final episode reward improvement. The DDPM-based implementation delivers a similar level of performance with lower volatility.
In contrast, the VAE-based framework attains comparable final reward to the non-pretrained baseline without convergence acceleration, despite a $53\%$ initial reward advantage. Notably, the GAN-based approach fails catastrophically, exhibiting $14.6\%$ final reward degradation, which underscores the critical importance of accurate state transition modeling in CMDP-based pretraining architectures. 
Furthermore, the performance of the Lyapunov baseline is approximately equivalent to that of the PPO agent trained for $300$ episodes. Both the VAE-ChMDN-based and DDPM-based pretrained agents outperform this level.

The simulation results show that quality of virtual CMDP modeling directly impacts pretraining performance. Our proposed pretraining framework, which leverages KAN for reward and cost prediction, VAE-ChMDN for initial-state distribution modeling, and EA-CGMM algorithm for state transition modeling, constructs an accurate virtual CMDP through offline transition tuples with acceptable computational complexity. This enables efficient DRL pretraining for the delay violation rate-constrained CF-MIMO resource allocation scenario.

\subsection{Scalability Analysis of Proposed Pretraining Framework}
\label{Scability}
In this subsection, we evaluate the scalability of the proposed pretraining framework in terms of both computational complexity and performance as the system scale increases.
To this end, we design two sets of experiments. First, to assess computational complexity, the system parameters $(B, U, K, M)$ are progressively scaled from $(2,2,2,2)$ to $(8,8,8,8)$, and the inference FLOPs of the PPO policy and the three modules in the virtual CMDP are measured. The hidden layer sizes of all modules are kept identical to those in previous experiments to ensure a fair comparison. 
Second, for performance evaluation, a larger system configuration with $(B, U, K, M) = (4,6,6,6)$ is considered. To ensure feasibility of the delay violation constraint and to facilitate more diverse transition tuple collection under the random behavior policy,
the maximum transmit power $P_{\max}$ is increased to $30\,\mathrm{dBm}$, and the delay violation rate threshold is set to $d=2\%$.
\begin{figure}
	\centering
	\includegraphics[width=3.3in]{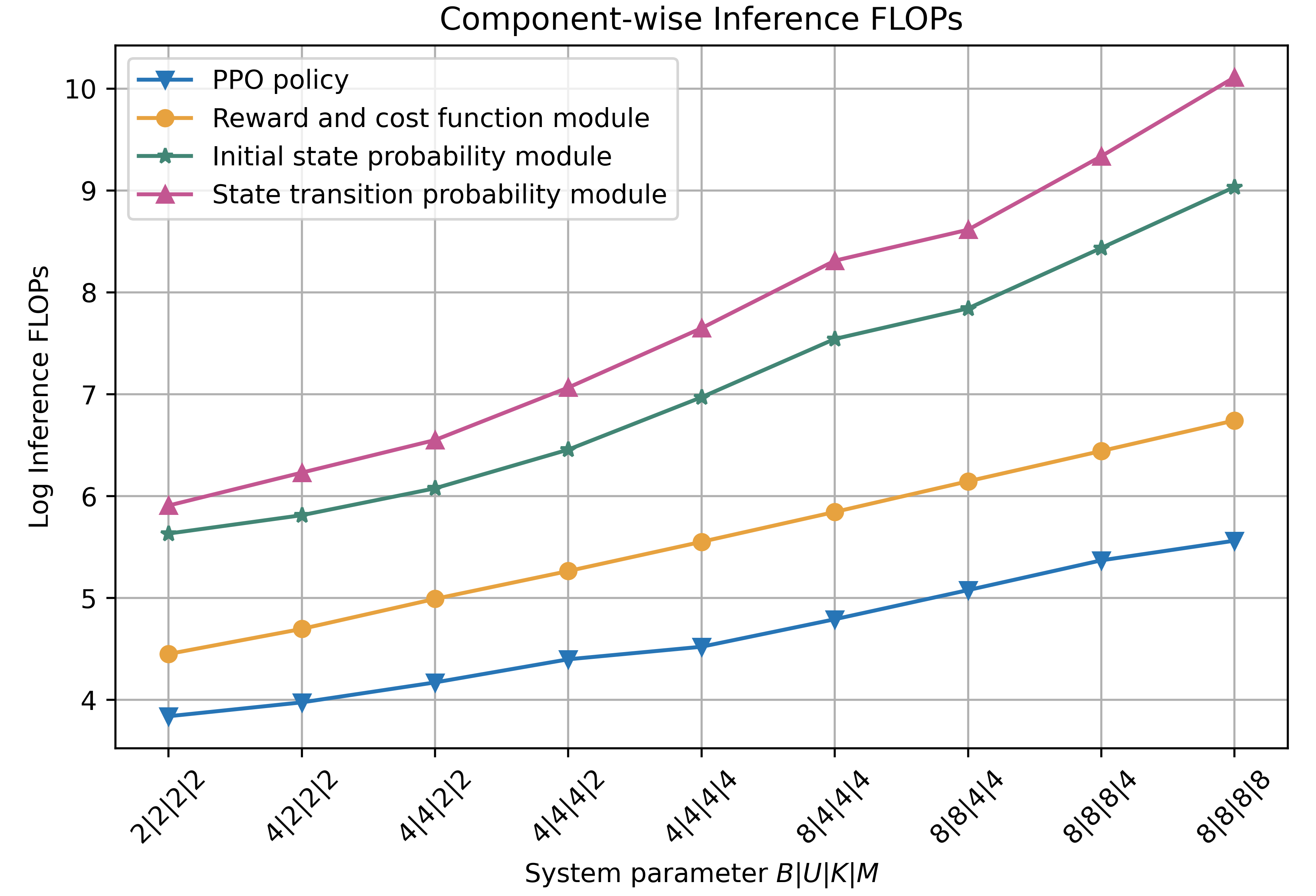}
	\caption{Log inference FLOPs of each module versus system scale $(B, U, K, M)$.}
	\label{Scalfigure_flops}
\end{figure}
\begin{figure}
	\centering
	\includegraphics[width=3.3in]{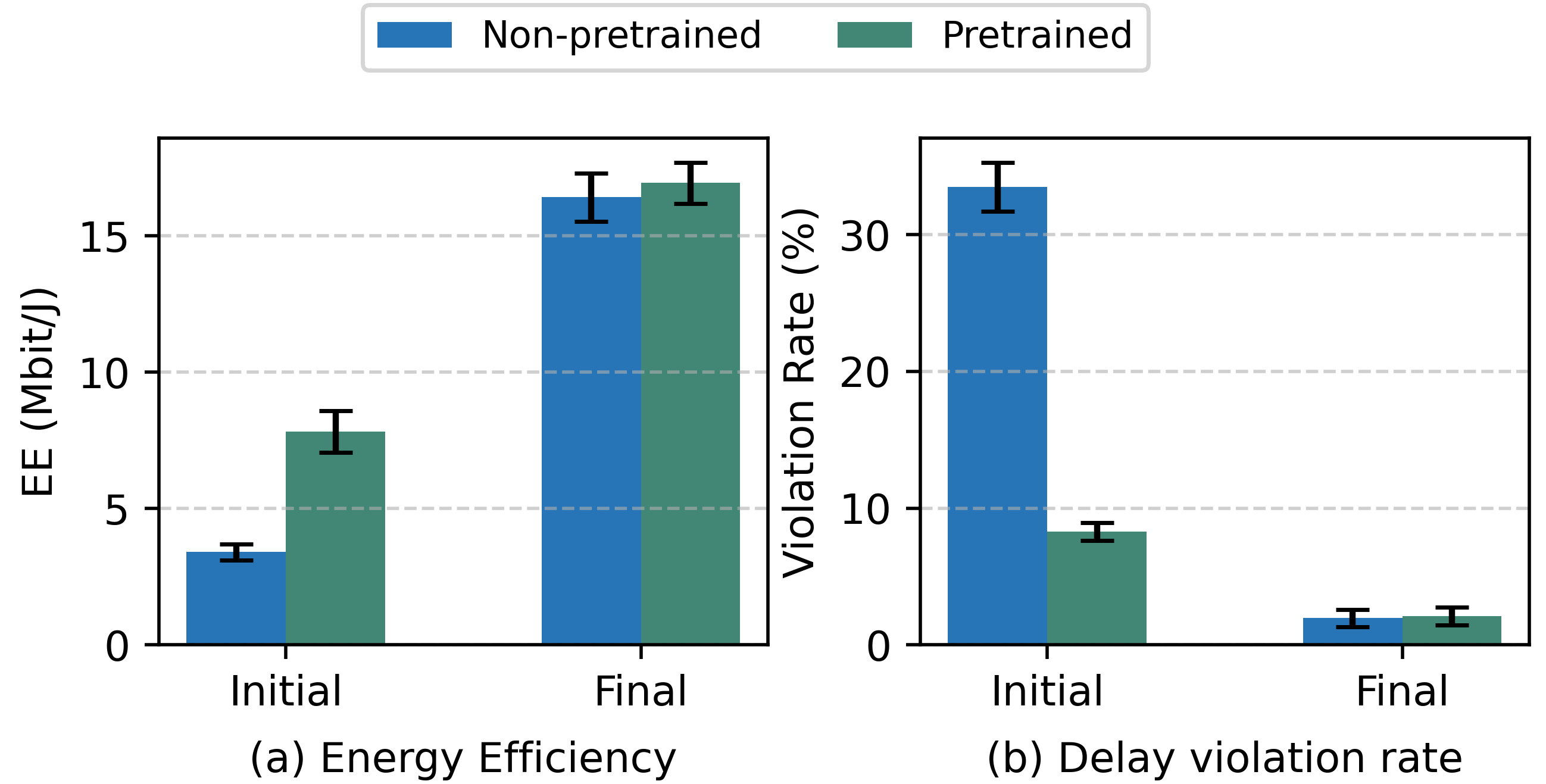}
	\caption{Performance comparison of pretrained and non-pretrained PPO policies in a CF-MIMO system with $(B, U, K, M) = (4,6,6,6)$. Error bars indicate the standard deviation over $8$ random seeds.}
	\label{Scalfigure_ee_c}
\end{figure}

Fig.~\ref{Scalfigure_flops} presents the component-wise computational complexities. The state transition probability module and the
initial state probability module, both based on VAE-MDN, consistently incur relatively high inference FLOPs. In contrast, the KAN-based reward and cost function module and the MLP-based PPO policy remain computationally lightweight.
Moreover, since the channel state information can be processed independently for each user, the FLOPs growth of the primary operational modules is less sensitive to the increase in $U$ than to the scaling of $B$, $K$, and $M$. This indicates that the proposed framework exhibits favorable scalability with respect to the number of users in CF-MIMO systems.

The policy performance is illustrated in Fig.~\ref{Scalfigure_ee_c}. Compared with non-pretrained baseline, the pretrained PPO achieves an initial EE improvement of $4.4$ Mbit/J and reduces the delay violation rate by $25\%$, while maintaining comparable final performance. Furthermore, it converges after approximately $1800$ training episodes, whereas the baseline requires around $2500$ episodes, corresponding to a $28\%$ reduction in training overhead.
Compared with the fine-tuning results in Section~\ref{Per_pretraining}, the gains in convergence speed and final performance decrease as the system scale increases. This degradation can be attributed to the reduced robustness of the constructed virtual CMDP and the increased training difficulty of PPO.

Future work will investigate incorporating the sparsity of channel covariance matrix into the Cholesky factor decomposition of VAE-ChMDN to reduce model complexity and improve the fidelity of the constructed virtual CMDP, thereby enhancing the scalability of the proposed pretraining framework.

\section{Conclusion}
\label{Conclusion}
This paper has proposed a virtual CMDP-based offline pretraining framework for the delay violation rate-constrained resource allocation problem in CF-MIMO downlink systems.
The framework utilizes KAN for reward and cost prediction, VAE-ChMDN for initial-state distribution modeling and the proposed EA-CGMM algorithm for state transition modeling. Notably, the proposed EA-CGMM approach employs the explicit GMM modeling for state transition tuples to analytically infer the next-state distribution, mitigating the data sparsity and distribution shift issues. 
Simulation results have shown that the KAN achieved lower prediction MAEs than MLP while maintaining approximately unbiased estimation. VAE-ChMDN and EA-CGMM attained comparable performance to SOTA DDPM and DDIM with classifier-free guidance, respectively, while reducing computational complexity significantly. Compared to the non-pretrained agent, the pretrained one achieved twice the initial EE with a low delay constraint violation rate of $1\%$, ultimately converging to a $4.7\%$ superior EE with $50\%$ fewer exploration steps.
These findings have collectively validated the accuracy of virtual CMDP modeling and the effectiveness of the proposed pretraining framework. 

In future work, several system-level extensions will be investigated, including the impact of system load and fronthaul capacity constraints on the latency violation probability in CF-MIMO systems, 
as well as more strict per-UE QoS satisfaction and proportional fairness-type optimization objectives.
From a methodological perspective, incorporating the sparsity of the channel covariance matrix into the Cholesky decomposition of the VAE-ChMDN can be further explored to reduce model complexity and improve the fidelity of the constructed virtual CMDP, thereby enhancing the scalability of the proposed pretraining framework.

\section*{Acknowledgments}
The authors would like to thank Dusit Niyato (Nanyang Technological University) for helpful discussions and thoughtful comments on the manuscript.

{\appendix[Derivations of conditional and marginal distributions in the multivariate Gaussian case]\label{appendix}
	An $n$-dimensional multivariate Gaussian distribution $\boldsymbol{x} \sim \mathcal{N}(\boldsymbol{\mu},\boldsymbol{\Sigma})$ can be parameterized by the mean vector $\boldsymbol{\mu} \in \mathbb{R}^n$ and the Cholesky factor $ \boldsymbol{U} \in \mathbb{R}^{n \times n}$. The matrix $\boldsymbol{U}$ is an upper triangular matrix with positive diagonal elements and the precision matrix satisfies $\boldsymbol{\Sigma}^{-1} = \boldsymbol{U}^\top\boldsymbol{U}$. Sampling is performed as follows:  
	\begin{equation}
	\boldsymbol{x} = \boldsymbol{U}^{-1}\boldsymbol{\epsilon} + \boldsymbol{\mu}, \quad \boldsymbol{\epsilon} \sim \mathcal{N}(\boldsymbol{0}, \boldsymbol{I}).
	\end{equation} 
	Partitioning $\boldsymbol{x} \) into \( \boldsymbol{x}_1 \in \mathbb{R}^m$ and $\boldsymbol{x}_2 \in \mathbb{R}^{n-m}$, the joint sampling process is:  
	\begin{equation}\label{DecomposeSample}
		\begin{bmatrix}
			\boldsymbol{x}_1 \\ \boldsymbol{x}_2
		\end{bmatrix}
		=
		\begin{bmatrix}
			\boldsymbol{U}_{11}^{-1} & -\boldsymbol{U}_{11}^{-1}\boldsymbol{U}_{12}\boldsymbol{U}_{22}^{-1} \\ 
			\boldsymbol{0}             & \boldsymbol{U}_{22}^{-1}
		\end{bmatrix}
		\begin{bmatrix}
			\boldsymbol{\epsilon}_1 \\ \boldsymbol{\epsilon}_2
		\end{bmatrix}
		+
		\begin{bmatrix}
			\boldsymbol{\mu}_1 \\ \boldsymbol{\mu}_2
		\end{bmatrix},
	\end{equation}  
	where $\boldsymbol{U}_{11} \in \mathbb{R}^{m \times m}$, $\boldsymbol{U}_{12} \in \mathbb{R}^{m \times (n-m)}$, and $\boldsymbol{U}_{22} \in \mathbb{R}^{(n-m) \times (n-m)}$ are submatrices of $\boldsymbol{U}$, formulated as:
	\begin{equation}
		\boldsymbol{U}=
		\begin{bmatrix}
			\boldsymbol{U}_{11} & \boldsymbol{U}_{12} \\
			\boldsymbol{0} & \boldsymbol{U}_{22}
		\end{bmatrix}.
	\end{equation}
	
	
	From \eqref{DecomposeSample}, the marginal distribution of $ \boldsymbol{x}_2$ is:  
	\begin{equation}\label{aeq:margin}
	\boldsymbol{x}_2 = \boldsymbol{U}_{22}^{-1}\boldsymbol{\epsilon}_2 + \boldsymbol{\mu}_2 \quad \Rightarrow \quad \boldsymbol{x}_2 \sim \mathcal{N}(\boldsymbol{\mu}_2, \boldsymbol{\Sigma}_2),
	\end{equation}  
	with precision matrix $\boldsymbol{\Sigma}_2^{-1} = \boldsymbol{U}_{22}^\top\boldsymbol{U}_{22}$.
	
	Given specific $\boldsymbol{x}_2^*$, the conditional distribution $p\left(\boldsymbol{x}_1 \mid \boldsymbol{x}_2^*\right)$ is derived as follows:
	
	\subsubsection{Conditional Mean Derivation}\label{app:cond_mean}  
	Substituting $\boldsymbol{\epsilon}_2^* = \boldsymbol{U}_{22}(\boldsymbol{x}_2^* - \boldsymbol{\mu}_2)$ into \eqref{DecomposeSample}:  
	\begin{equation}\label{aeq:cori}
		\boldsymbol{x}_1 = \boldsymbol{U}_{11}^{-1}\boldsymbol{\epsilon}_1 + \left( \boldsymbol{\mu}_1 - \boldsymbol{U}_{11}^{-1}\boldsymbol{U}_{12}(\boldsymbol{x}_2^* - \boldsymbol{\mu}_2) \right).
	\end{equation}
	According to $\boldsymbol{\epsilon}_1 \sim \mathcal{N}(\boldsymbol{0}, \boldsymbol{I})$, the mean vector of \( \boldsymbol{x}_1 \mid \boldsymbol{x}_2^* \) is:
	\begin{equation}\label{aeq:cmean}
	\boldsymbol{\mu}_{1\mid2} = \boldsymbol{\mu}_1 - \boldsymbol{U}_{11}^{-1}\boldsymbol{U}_{12}(\boldsymbol{x}_2^* - \boldsymbol{\mu}_2).
	\end{equation}
	
	\subsubsection{Conditional Precision Derivation}\label{app:cond_precision}  
	The precision matrix of $\boldsymbol{x}_1 \mid \boldsymbol{x}_2^*$ is:  
	\begin{equation}\label{aeq:ccov}
	\boldsymbol{\Sigma}_{1\mid2}^{-1} = \boldsymbol{U}_{11}^\top\boldsymbol{U}_{11}.
	\end{equation}
	Thus, the conditional distribution is:  
	\begin{equation}\label{aeq:cond}
	\boldsymbol{x}_1 \mid \boldsymbol{x}_2^* \sim \mathcal{N}\left( \boldsymbol{\mu}_1 - \boldsymbol{U}_{11}^{-1}\boldsymbol{U}_{12}(\boldsymbol{x}_2^* - \boldsymbol{\mu}_2), \, \left( \boldsymbol{U}_{11}^\top\boldsymbol{U}_{11} \right)^{-1} \right).
	\end{equation}}

\bibliographystyle{IEEEtran}
\bibliography{references}  

\end{document}